\documentclass[sigconf]{acmart}
\usepackage{balance}
\usepackage{stfloats}

\AtBeginDocument{%
  }

\copyrightyear{2026}
\acmYear{2026}
\setcopyright{cc}
\setcctype{by-nc-nd}
\acmConference[CHI '26]{Proceedings of the 2026 CHI Conference on Human Factors in Computing Systems}{April 13--17, 2026}{Barcelona, Spain}
\acmBooktitle{Proceedings of the 2026 CHI Conference on Human Factors in Computing Systems (CHI '26), April 13--17, 2026, Barcelona, Spain}
\acmPrice{}
\acmDOI{10.1145/3772318.3790817}
\acmISBN{979-8-4007-2278-3/2026/04}

\begin{document}

\title[Generative Muscle Stimulation: Physical Assistance by Constraining Multimodal-AI with Embodied Knowledge]{Generative Muscle Stimulation: Providing Users with Physical Assistance by Constraining Multimodal-AI with Embodied Knowledge}

\author{Yun Ho}
\authornote{These authors contributed equally and are ordered alphabetically.}
\affiliation{
  \institution{University of Chicago}
  \country{United States}
}
\email{yunho@uchicago.edu}

\author{Romain Nith}
\authornotemark[1]
\affiliation{
  \institution{University of Chicago}
  \country{United States}
}
\email{rnith@uchicago.edu}

\author{Peili Jiang}
\affiliation{
  \institution{University of Chicago}
  \country{United States}
}
\email{peilij@uchicago.edu}

\author{Steven He}
\affiliation{
  \institution{University of Chicago}
  \country{United States}
}
\email{stevenhestudent@gmail.com}

\author{Bruno Felalaga}
\affiliation{
  \institution{University of Chicago}
  \country{United States}
}
\email{brunofelalaga@uchicago.edu}

\author{Shan-Yuan Teng}
\affiliation{
  \institution{University of Chicago}
  \country{United States}
}
\email{tengshanyuan@uchicago.edu}

\author{Rhea Seeralan}
\affiliation{
  \institution{University of Chicago}
  \country{United States}
}
\email{rheaseeralan@gmail.com}

\author{Pedro Lopes}
\affiliation{
  \institution{University of Chicago}
  \country{United States}
}
\email{pedrolopes@uchicago.edu}

\renewcommand{\shortauthors}{Ho and Nith et al.}

\begin{abstract}
Electrical muscle stimulation (EMS) can support physical-assistance (e.g., shaking a spray-can before painting). However, EMS-assistance is highly-specialized because it is (1) fixed (e.g., one program for shaking spray-cans, another for opening windows); and (2) non-contextual (e.g., a spray-can for cooking dispenses cooking-oil, not paint—shaking it is unnecessary). Instead, we explore a different approach where muscle-stimulation instructions are generated considering the user’s context (e.g., pose, location, surroundings). The resulting system is more general—enabling unprecedented EMS interactions (e.g., opening a pill bottle) yet also replicating existing systems (e.g., \emph{Affordance++}) without task-specific programming. It uses computer-vision/large-language-models to generate EMS-instructions, constraining these to a muscle-stimulation knowledge-base \& joint-limits. In our user-study, we found participants successfully completed physical-tasks while guided by generative-EMS, even when EMS-instructions were (purposely) erroneous. Participants understood generated gestures and, even during forced-errors, understood partial-instructions, identified errors, and re-prompted the system. We believe our concept marks a shift toward more general-purpose EMS-interfaces.
\end{abstract}



\begin{teaserfigure}
  \vspace*{-0.3cm}
  \includegraphics[width=\columnwidth]{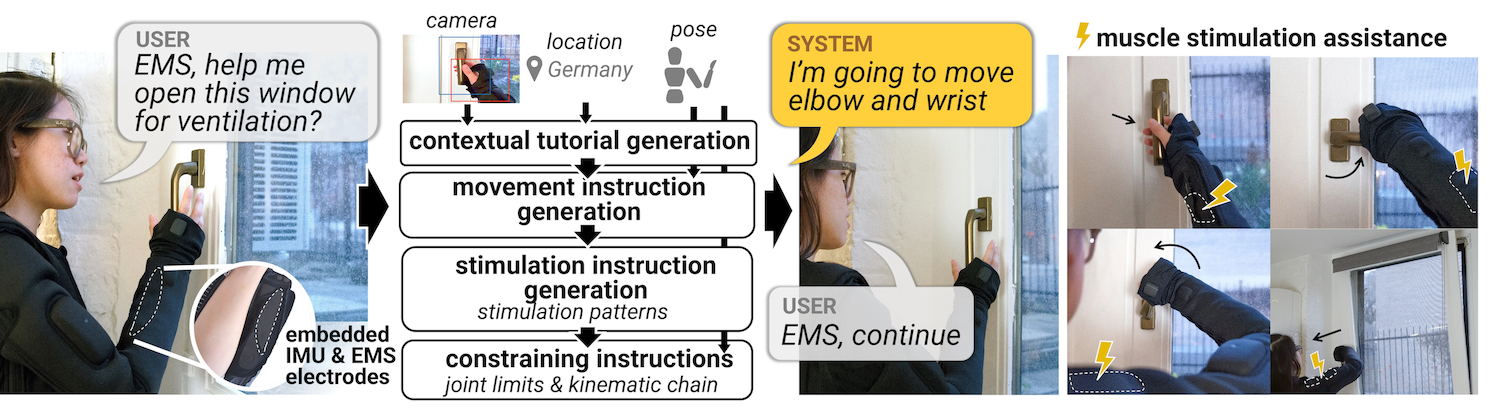}
  \vspace*{-0.8cm}
  \caption{Our system assists users via electrical muscle stimulation; except, it does not deliver fixed-instructions but generates them. This user requests help to open a window in a foreign country. Our system gathers information: body pose (IMU suit) + point-of-view (objects, hand-poses) + location (e.g., “Germany”), etc. These are fed to a multimodal-AI that creates textual-instructions—allowing it to infer that this window opens vertically. As AI-models have no information to respect joint-limits of the user’s body, our system constrains instructions with embodied knowledge. Thus, opening this window (used in our user study).}
  \Description{The figure depicts a pipeline from left to right. Left most: a user wearing a black suit hovers her hand on the handle of a window. A callout from her mouth states the request from the user. A call out from the sleeve shows embedded IMU \& EMS. Next to it, boxes indicating the processes of the system are shown. Right section shows the user asks for help from our system by stating "EMS help me open this window for ventilation" and the image depicts EMS-generated poses that assist the user into physically opening such a window—note the window is not a traditional window, it actually opens vertically (European-style window)}
  \label{fig:1}
\end{teaserfigure}

\maketitle

\vspace*{-0.4cm}
\section{Introduction}

\emph{Procedural}-\emph{knowledge} is denoted as the "know-how" of
performing a task, in contrast to knowing
facts~\cite{tenbergeProceduralDeclarativeKnowledge1999,cohenPreservedLearningRetention1980};
it includes the \emph{sequences of movements} needed to achieve the
task. For instance, knowing the mechanism of a pickle jar is not
synonymous with being able to open it effortlessly (it requires a
specific gesture and force, e.g., leveraging grip friction to enable a
strong twist). Supporting users in these situations is very challenging,
as procedural-knowledge is often acquired from lived-experiences, and
simply relaying facts is often insufficient
\cite{lewickiAcquisitionProceduralKnowledge1988,nishidaDigituSyncDualUserPassive2022,zhouEffectHapticFeedback2012}.

To tackle this, much effort has been dedicated to engineering interfaces
for physical assistance via \emph{force-feedback}
\cite{chenAssistingRepellingForceFeedback2013,guptaDesignControlPerformance2008,lopesAffordanceAllowingObjects2015,nakagakiLineFORMActuatedCurve2015,nithSplitBodyReducingMental2024,dupasquierHaptiknitDistributedStiffness2024,shenKinethreadsSoftFullBody2025,takahashiNovelSoftExoskeleton2019}.
One interface that has gained popularity in the last two decades, with
\textasciitilde150 publications on the topic in HCI alone
\cite{faltaousPerceptionActionReview2022}, is
electrical muscle stimulation (EMS) due to its wearability. Interactive
systems based on EMS move the user's body via electrically-actuated
movements. Examples of interactive EMS-assistance include: sign-language
\cite{nithDextrEMSIncreasingDexterity2021}, eyes-free navigation
\cite{takahashiCanSmartwatchMove2024}, piano-playing
\cite{niijimaMotorSkillDownloadSystemUsing2024,takahashiIncreasingElectricalMuscle2021},
and demonstrating movement instructions to assist users in operating
objects that users have not used before
\cite{lopesAffordanceAllowingObjects2015}---in
\emph{Affordance++}~\cite{lopesAffordanceAllowingObjects2015}, EMS is
used to help users discover actions, e.g., force users to shake a
spray-can before painting.

Unfortunately, the \textasciitilde150 systems in HCI using EMS for
physical assistance \cite{faltaousPerceptionActionReview2022} are
highly-specialized due to their \textbf{(1) rigid programming:} movement
instructions are crafted in advance rather than generated in real-time
(e.g., \emph{Affordance++} \cite{lopesAffordanceAllowingObjects2015}
provides only six fixed EMS-instruction sequences) resulting in limited
sets of predefined movements; and \textbf{(2) non-contextual}:
EMS-systems ignore contextual-cues that could enable assisting users in
more complex situations (e.g., \emph{Affordance++} shakes a spray-can
for painting, but it would fail to recognize that a cooking oil
spray-can does not require shaking).

Towards our goal of advancing the research in interactive systems based
on EMS, we explored a new avenue: we engineered a system that
\emph{generates} EMS-instructions given the user's request \& context.
As depicted in \hyperref[fig:1]{Figure 1}, our system takes
advantage of multimodal-AI: it takes in a user's spoken-request and
images from their point-of-view (POV); it uses computer-vision (e.g.,
detect objects/hands) large-language-models (e.g., reason about objects,
situations, surroundings), EMS-knowledge (e.g., feasible movements) and
joint-information (e.g., joint-limits, kinematic-chain) to
\emph{generate} EMS\emph{-}instructions that were \emph{not} part of the
system's built-in programming. In this particular example, where the
user is facing a tilt-turn window (common in parts of Europe) that can
be opened by turning the handle vertically, our EMS-system generated EMS
movements that actuated the user's body, depicting the movements needed
to open the window (i.e., stimulate the shoulder to turn the handle,
then biceps to pull).

\begin{figure*}[h!]
\centering
  \vspace*{-0.3cm}
  \includegraphics[width=\textwidth]{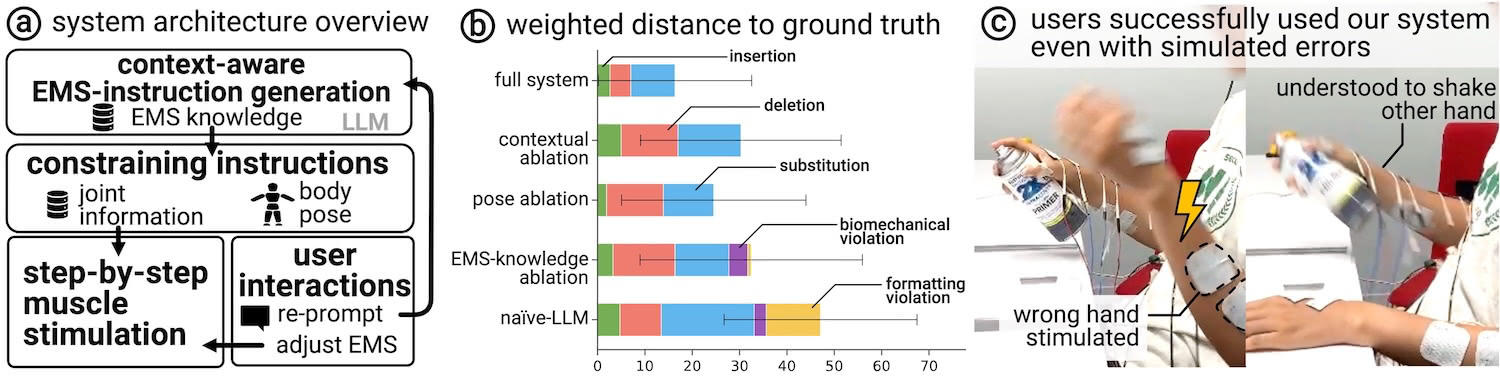}
  \vspace*{-0.6cm}
  \caption{Preview of contributions: (a) our system architecture; (b) ablation study suggests it produced \textasciitilde3x fewer errors; (c) results from user study suggest they were able to use our system to perform physical tasks, even if instructions contained simulated errors.}
  \Description{(a) A system architecture overview is shown containing blocks of subsystems. (b) A plot shows the comparison of errors from various techniques. (c) Photos show that a person with electrodes on their hands, attempting to shake a spray bottle}
  \vspace*{-0.4cm}
  \label{fig:2}
\end{figure*}

\vspace*{-0.2cm}
\subsection{Scope and contribution}\label{scope-and-contribution}

Our contribution is two-fold, one \textit{technical} and one \textit{conceptual}.

Our conceptual contribution is proposing a type of \textit{physical} assistance based on multimodal-AI. Unlike traditional AI-based interfaces, the feedback that our AI provides to its users is primarily in the form of (electrically actuated) \textit{muscle movements}. This type of unique AI-feedback allowed us to put forward a novel form of \textbf{embodied-AI}: an AI that understands \& assists with body movements, rather than with language.

Our technical contribution is improving EMS-based interactive systems by leveraging multimodal-AI reasoning to \textit{generate} electrical muscle stimulation instructions that were \emph{not} part of the system's built-in programming. As such, our
work builds on EMS research \cite{faltaousPerceptionActionReview2022}
and contributes a generative-EMS architecture that yields new benefits
and insights into EMS research. As depicted in
\hyperref[fig:2]{Figure 2}, we complement our (a) open-source
implementation which we hope will accelerate new research in this area
with (b) an ablation study that characterized our implementation; and
(c) a user study, where we found that, even when such systems fail
(e.g., generated incorrect instructions), participants were able to
recover and interpret gestures and successfully perform the intended
tasks.

Finally, it is important to underscore that our contribution is centered
only around proposing, implementing, and measuring the generation of EMS
instructions rather than comparing EMS-based assistance against other
modalities (e.g., visuals, auditory), or improving upon the existing
limitations of EMS (e.g., its practicality). Moreover, while there are
many technical routes to implement our concept (e.g., other
multimodal-AI techniques, reinforcement learning, simulators), our key
contribution is not to exhaustively implement all these alternatives, but
to examine the value of instruction generation \emph{for interactive
EMS-systems}.

\vspace*{-0.2cm}
\section{Related work}\label{related-work}

Our work is focused on advancing interactive systems based on EMS. While
EMS is not the only actuator capable of physical assistance, we focus on
it due to its wearability when compared to mechanical actuators
\cite{carignanDevelopmentExoskeletonHaptic2009,choiSoftControllableHigh2018,choiWolverineWearableHaptic2016,guDexmoInexpensiveLightweight2016,hinchetDextrESWearableHaptic2018,nagaiWearable6DoFWrist2015,onenDesignActuatorSelection2014}.
As we aim to generate EMS-instructions, we review key works in
multimodal-AI. Given the wide range of this area, we provide only a
succinct review---refer to
\cite{baltrusaitisMultimodalMachineLearning2019,dongNextGenerationIntelligentAssistants2023,voulodimosDeepLearningComputer2018,wangDeep3DHuman2021a,zouReviewObjectDetection2019}
for extensive reviews.

\vspace*{-0.2cm}
\subsection{\texorpdfstring{Conveying \emph{procedural-knowledge}
through haptic
feedback}{Conveying procedural-knowledge through haptic feedback}}\label{conveying-procedural-knowledge-through-haptic-feedback}

\emph{Procedural}-\emph{knowledge} is denoted as the "know-how" of
performing a task
\cite{tenbergeProceduralDeclarativeKnowledge1999,cohenPreservedLearningRetention1980}.
It supports the acquisition/retention of manual tasks, and it does not
depend just on declarative knowledge or vision
\cite{johnsonProceduralMemorySkill2003,pinzonSkillLearningKinesthetic2017,willinghamDevelopmentProceduralKnowledge1989}.
Studies found that procedural-knowledge is learned implicitly via
experience---people can learn it even if they cannot articulate
it~\cite{lewickiAcquisitionProceduralKnowledge1988}---physical know-how
is usually shown or felt rather than described
\cite{nishidaDigituSyncDualUserPassive2022}. People tend to \emph{learn
by doing}---hands-on experiences involving active manipulation allow
them to associate movement instructions and more easily remember the
task
\cite{franchakLearningDoingAction2010,mccarthyHowDoesAssembling2024,minogueHapticsEducationExploring2006}.

One promising way to deliver a hands-on user experience is through
\emph{haptic feedback}. For instance, providing movement guidance with
haptic feedback benefits learning physical tasks, such as surgical
procedures
\cite{oquendoHapticGuidanceHaptic2024,zhouEffectHapticFeedback2012},
sports~\cite{renEnhancingMotorSkills2025,tayalVMLHSTDevelopmentEfficient2023},
welding \cite{zhongAutomatedKinestheticTrainer2013}, as well as
memorizing movement sequences
\cite{choudharyAdaptiveElectricalMuscle2025,nishidaDigituSyncDualUserPassive2022}.
In fact, several studies confirmed the added value of haptic feedback
with respect to motor learning
\cite{grantAudiohapticFeedbackEnhances2019,marchal-crespoEffectHapticGuidance2013,morrisHapticFeedbackEnhances2007,sigristAugmentedVisualAuditory2013}.
With respect to the scope of our work---i.e., a form of haptics known as
electrical muscle stimulation---many systems demonstrated the benefits
of using EMS to actuate the user's body to demonstrate procedural
knowledge
\cite{faltaousEMStrikerPotentialsEnhancing2022,faltaousGeniePuttAugmentingHuman2021,hassanFootStrikerEMSbasedFoot2017,lopesAffordanceAllowingObjects2015,tatsunoSupportiveTrainingSystem2017a}.
In fact, recent findings suggest that EMS can, in certain contexts,
outperform other types of haptic feedback for skill transfer
\cite{leeHapticusExploringEffects2025}. It is important to underscore
that many other modalities can be used to convey procedural knowledge
(e.g., visuals, audio); however, these lay outside the scope of our
paper, which is focused on \emph{electrical muscle  stimulation}.

\subsection{\texorpdfstring{Electrical muscle stimulation (EMS) as force
feedback
}{Electrical muscle stimulation (EMS) as force feedback }}\label{electrical-muscle-stimulation-ems-as-force-feedback}

EMS actuates the body by electrically contracting muscles. While it
originated in
rehabilitation~\cite{strojnikProgrammedSixChannelElectrical1979}, it
became a popular choice for force-feedback, as its hardware is smaller
than mechanical-actuators~\cite{lopesImmensePowerTiny2017}. This led to \textasciitilde150
EMS-publications in HCI \cite{faltaousPerceptionActionReview2022}, such
as AR/VR
\cite{chengPairedEMSEnhancingElectrical2024,kimEffectMultisensoryPseudoHaptic2022,lopesImpactoSimulatingPhysical2015,lopesProvidingHapticsWalls2017},
or skill-acquisition
\cite{faltaousGeniePuttAugmentingHuman2021,hassanFootStrikerEMSbasedFoot2017,lopesAffordanceAllowingObjects2015,tatsunoSupportiveTrainingSystem2017a}.
Given our scope, we focus on assistance-based EMS-systems (see
\cite{faltaousPerceptionActionReview2022} for a more general discussion of application areas in interactive-EMS).

Typically, assistance-based EMS-systems convey information by moving the
body. For instance, \emph{Affordance++}
\cite{lopesAffordanceAllowingObjects2015} provides EMS movements for six
objects the user might encounter (e.g., movements to shake a spray-can
before painting). However, these EMS-instructions are fixed---when a
user encounters a new object that requires instructions not found in the
system's built-in programming (e.g., open a pill bottle),
current generation EMS-systems cannot adjust their movements to tackle
this. Similarly, the instructions delivered by traditional EMS-systems
are non-contextual. For instance, while \emph{Affordance++}
\cite{lopesAffordanceAllowingObjects2015} delivers EMS-instructions to
operate a spray-can while painting (e.g., it shakes the can), yet, these
would fail to assist a user who is spraying cooking oil (no shaking is
needed)---in this case, contextual cues could aid in resolving this by
including the visuals of the tools or surroundings (e.g., kitchen,
frying pan). These contextual cues need not be visual; they could be
based on the user's geolocation (e.g., some tools work differently in
different regions of the globe) and so forth.

Our research question is as follows: if EMS-systems were given the
ability to provide flexible instructions and understand contextual cues,
could it widen their scope? (e.g., support users even in situations not
explicitly included in their programming?). To embed this novel level of
understanding in EMS, we turned to a range of technical approaches,
namely, multimodal AI and constraining AI output with additional
domain-knowledge (e.g., EMS knowledge).

\subsection{\texorpdfstring{Multimodal-AI to reason on unknown
objects/situations
}{Multimodal-AI to reason on unknown objects/situations }}\label{multimodal-ai-to-reason-on-unknown-objectssituations}

Integrating large-language-models (LLMs) with computer-vision has
enabled reasoning on more complex visual contexts---leading to
vision-language-models (VLMs)
\cite{zhangVisionLanguageModelsVision2024,zhouLearningPromptVisionLanguage2022}.
Many models exist that can either reason on visual concepts learned from
natural language descriptions (e.g., \emph{CLIP}
\cite{radfordLearningTransferableVisual2021}) or hybridize
object-detection with language models (e.g.,
\cite{hanFewShotObjectDetection,zangContextualObjectDetection2025}). An
area of rapid progress is refining multimodal pipelines to improve
accuracy (e.g., \emph{InstructGPT}
\cite{ouyangTrainingLanguageModels2022} and \emph{Self-RAG}
\cite{asaiSelfRAGLearningRetrieve2023}--- to cite a few). Given the
breadth of multimodal-AI, we recommend detailed surveys
\cite{baltrusaitisMultimodalMachineLearning2019,huVisionBasedMultimodalInterfaces2025}.

Multimodal-AI was leveraged in many interfaces for assistance (an
exhaustive list of LLM applications in HCI is provided in
\cite{pangUnderstandingLLMificationCHI2025}), including assisting blind
users in recognizing objects
\cite{changWorldScribeContextAwareLive2024,leeGazePointARContextAwareMultimodal2024}
or guiding procedural tasks via audiovisual feedback
\cite{arakawaPrISMObserverInterventionAgent2024,huhVid2CoachTransformingHowTo2025,laiLEGOLearningEGOcentric2025}.
While audiovisual guidance is important, we focus on a different type of
assistance: movement assistance via force-feedback---i.e., demonstrating
a task not with audiovisuals but via \emph{movements}.

With respect to generating force-feedback instructions, we can draw
parallels to robotics. Multimodal-AI models, supplemented with
robotic-specific training or fine-tuning, have been shown to generate
task-planning for robotic manipulators
\cite{IntroducingGeminiRobotics2025,nodaMultimodalIntegrationLearning2014,wangLargeLanguageModels2025}.
We also use multimodal-AI to generate instructions, but we focus on
generating EMS-instructions to be delivered directly to the user's body
(not via an external robot), which requires enabling the system to
consider the user's current body pose, joint-limits, the capabilities of
EMS, etc.

\subsection{Balancing general-reasoning with
domain-specificity}\label{balancing-general-reasoning-with-domain-specificity}

Foundation models are suitable for reasoning on \emph{general tasks}
since, to some degree, they reason through abstract instructions
\cite{geOpenAGIWhenLLM2023}, can parse these into steps
\cite{laiLEGOLearningEGOcentric2025,meyerPotentialsLargeLanguage2024,wuAIChainsTransparent2022},
extrapolate to unseen examples
\cite{dorbalaCanEmbodiedAgent2024,kojimaLargeLanguageModels2022}, find
clues in text or images
\cite{zhangVinVLRevisitingVisual2021,zhangVisionLanguageModelsVision2024},
etc.---all of which are critical properties if a system desires to
assist users in a task that is \emph{not} part of its built-in
programming. This is the key reason why we leverage LLMs/VLMs to
generate instructions. However, as we will detail in \hyperref[implementation]{\emph{Implementation Study}}, foundation models are not designed to output
electrical muscle stimulation and can fail/hallucinate when selecting the correct
joint to move, request moving a joint past its biomechanical limits, and
so forth.

A popular approach to tackle this is to \emph{constrain} the output of
the model to generate domain-specific responses. This can be implemented
in a number of ways, such as knowledge-graphs
\cite{tanPathsoverGraphKnowledgeGraph2025,zhangWaySpecialistClosing2025},
filters
\cite{liuWeNeedStructured2024,yuanDistillingScriptKnowledge2023}, and so
forth. For instance, \emph{CoScript}
\cite{yuanDistillingScriptKnowledge2023} uses an LLM to generate
instructions for tasks but filters the outputs using a similarity
score, allowing it to pick options that best match constraints set in
the prompt (e.g., ``make a cake for diabetics'' deprioritizes using
sugar); \emph{Scaling Up and Distilling Down} uses success-filtering for
LLM-guided robot skill acquisition
\cite{haScalingDistillingLanguageGuided2023} (this filter learns from
successful outputs, prioritizing these); \emph{SHAPE-IT}
\cite{qianSHAPEITExploringTexttoShapeDisplay2024} used retrieval
augmented generation to map LLM textual-instructions to pin-based
display animations, etc. Inspired by these, we leverage constraints to
transform textual-instructions into muscle-stimulation instructions. In
our case, the constraints consider \emph{EMS-knowledge} (e.g., user's
pose, joint-limits, kinematics).

\section{Walkthrough}\label{walkthrough}

To illustrate the flexibility that our approach adds to
AI interactions, we demonstrate via a walkthrough. In this example, our system acts as a form of \textbf{embodied-AI}—a system that, rather than assisting its user with (audiovisual) textual instructions, assists its user by means of generated (electrically actuated) muscle movements. 

\begin{figure}[h!]
\centering
  \includegraphics[width=\columnwidth]{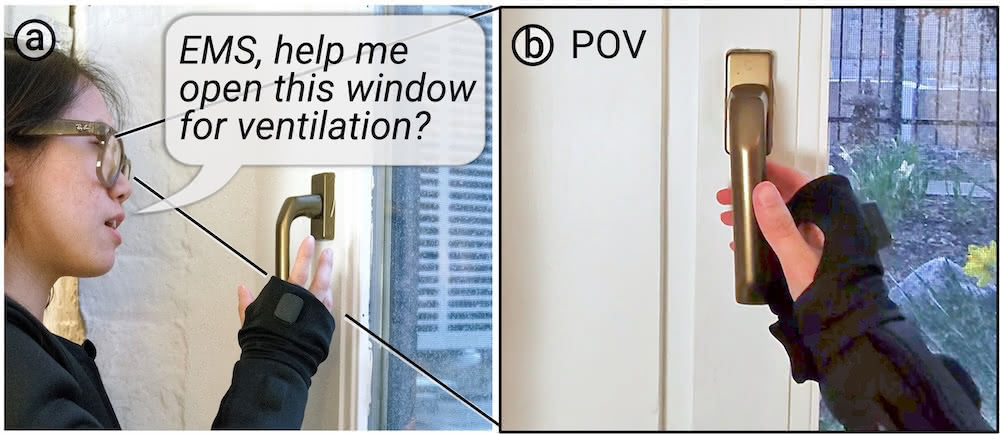}
  \vspace*{-0.6cm}
  \caption{(a) User asks for help with window. (b) Their POV.}
  \Description{A person holds the handle of the window, while a call out states "EMS, help me open this window for ventilation". A call out shows a picture from the first person's view.}
  \label{fig:3}
  \vspace*{-0.5cm}
\end{figure}

\hyperref[fig:3]{Figure 3} (a) depicts our user wearing a
rendition of our concept (see \hyperref[implementation]{\emph{Implementation Study}}), comprised of an electrical muscle stimulation \& motion-tracking suit (\emph{Teslasuit}), alongside
glasses with a camera and a microphone (\emph{Meta Ray-Ban}). Our user,
who is traveling to a foreign country, is struggling to operate objects.
Confused by a window, she asks our system, ``EMS, help me open this
window for ventilation?'' (``EMS'' is the wake-up word for invocation).

All
interactions in our examples depict unmodified EMS-outputs of
running our system as shown (e.g., using these voice prompts, these POV photos, electrodes placed as is, etc.). These two examples (opening a
non-traditional window and a pill bottle with a locking-mechanism) are
also featured in our studies (see \hyperref[technical-evaluation-via-an-ablation-study]{\emph{Ablation Study}} and 
\hyperref[user-study]{\emph{User Study}}.

\begin{figure}[h!]
\centering
  \includegraphics[width=\columnwidth]{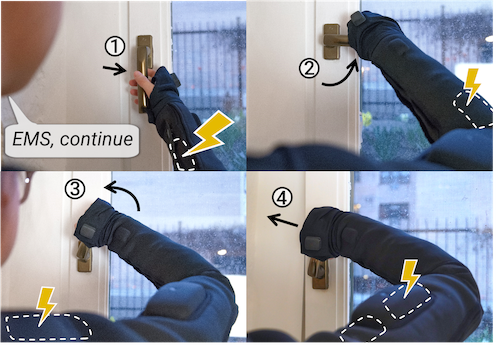}
  \vspace*{-0.6cm}
  \caption{EMS sequence that opens the window in tilt mode.}
  \Description{Step by step photos showing a person wearing suit, holding the window handle, turning it, and tilting the window.}
  \vspace*{-0.3cm}
  \label{fig:4}
\end{figure}

To assist the user, our system gathers information. It captures her
point of view and location (e.g., ``Germany'', the country she is
visiting). Using computer-vision, it extracts objects (``window'' and
``handle'') and hand poses. These are passed to our architecture, which
uses multimodal-AI and our EMS-knowledge to generate instructions (see
\hyperref[implementation]{\emph{Implementation Study}}). Now, as depicted in
\hyperref[fig:4]{Figure 4}, our system is ready-to-act and
informs the user (``I will move fingers and arm'') and \emph{waits for
confirmation}. After she confirms (``EMS, continue''), it delivers
movement instructions via EMS, causing their body to grasp the handle
(finger flexion), turn the handle to the upwards position (wrist
pronation \& shoulder abduction), and pull the handle to tilt the window
(elbow flexion).

\begin{figure}[h!]
\centering
  \includegraphics[width=\columnwidth]{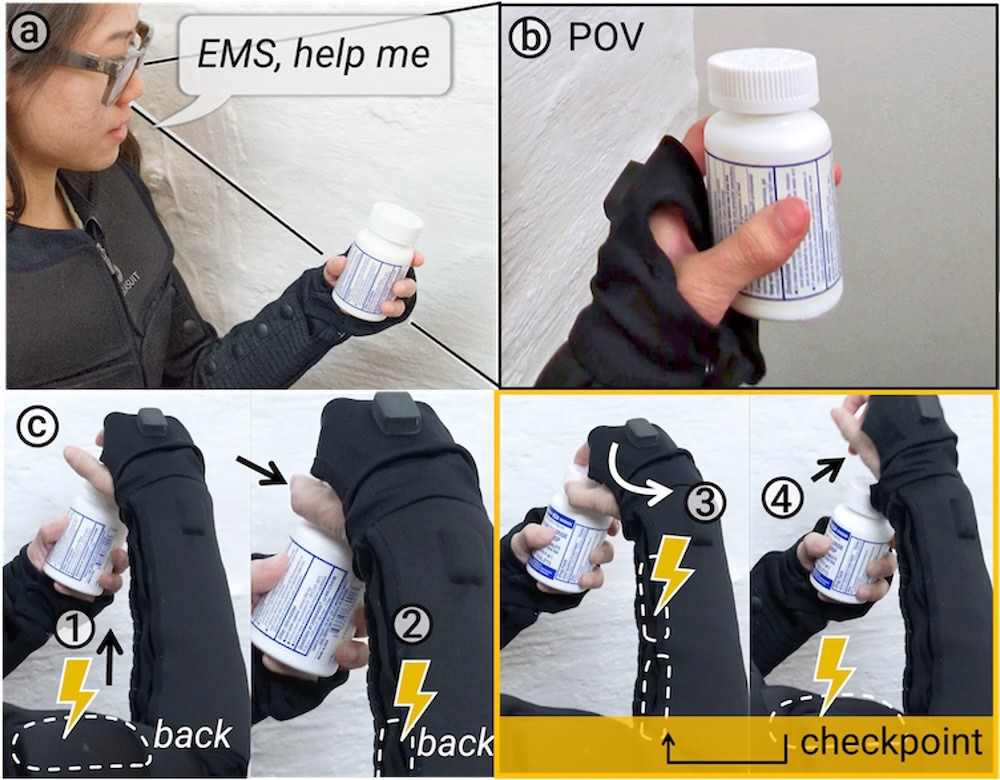}
  \vspace*{-0.6cm}
  \caption{(a) User asks for help with a pill bottle. (b) Their POV. (c) The system can understand some degree of the state of objects, which is used to add “checkpoints” to the muscle-stimulation instructions, e.g., repeat until the cap is removed. (Although, as we will see in the next step of this walkthrough, this will not open this type of pill bottle.)}
  \Description{A person wearing glasses and a glove asks for help while holding the pill bottle. Another part is shown holding a pill bottle from a first-person perspective. Right side shoes step by step photos showing a person's arm opening a pill bottle with electrical muscle stimulation on the arm. An arrow points from the last step to a previous step at checkpoint.}
  \label{fig:5}
  \vspace*{-0.4cm}
\end{figure}

After this, in \hyperref[fig:5]{Figure 5}, our user tries to
open a pill bottle they got at the pharmacy, featuring a
locking-mechanism unfamiliar to them. Failing to open it, she asks,
``EMS, help me''. This time, we opted for an example where the user's
request is vague (no mention of goal/object); yet our system gathers
contextual-cues to generate EMS-instructions.

\hyperref[fig:5]{Figure 5} (c) depicts that, after the user's
confirmation, our system delivers the EMS-instructions that cause her
hands to squeeze the cap (finger flexion) and turn the cap (wrist
abduction). During EMS, our system uses checkpoints to select the next
EMS-instruction as needed. For instance, when not seeing that the cap
was removed, it adjusts the EMS plan to open the cap in
\emph{repetitive} movements (grasp fingers, turn clockwise, release
fingers, turn counter-clockwise---repeat). It executes this until it
reaches the checkpoint (i.e., determines the cap is out) or the user
verbally stops the interaction (e.g., ``EMS stop'') or the user moves
against the EMS feedback (e.g., employed also by
\cite{lopesAffordanceAllowingObjects2015} as a quick-stop).

\begin{figure}[h!]
\centering
  \vspace*{-0.3cm}
  \includegraphics[width=\columnwidth]{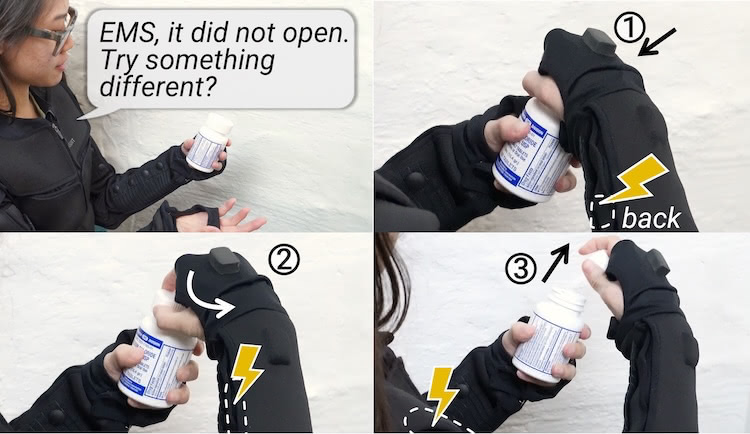}
  \vspace*{-0.6cm}
  \caption{(a) User intervenes and asks for a different approach, which triggers our system to generate a new sequence of electrical muscle stimulation assistance.}
  \Description{A person wearing glasses and a glove asks for help to try a different way. Step by step photos show both hands and a pill bottle. The hands are pointed with a lightning label with arrows showing the directions.}
  \label{fig:6}
\end{figure}

However, note that we intentionally complicated this example to illustrate a
failure-case and how the added flexibility might allow our system to
recover. Despite the EMS-assistance in the last step, the pill bottle did not open. Thus, not seeing the cap removed, the user intervenes: ``EMS, it did not
open. Try something different?''. In response, our system restarts, but appends the
output of the previous interaction. This allows the reasoning to
generate another solution, which is depicted in \hyperref[fig:6]{Figure
6} (``I can push fingers and turn wrist''). The user confirms, and the
EMS actuates the body, which causes the bottle cap to open---turns out
it was a \emph{push-turn} cap mechanism rather than a
\emph{squeeze-turn}.

Finally, we purposely illustrated extremely verbose examples in which the muscle stimulation
plan is read aloud to the user via text-to-speech. This is not the only way such an embodied-AI system can preview its muscle-based instructions to the user. In fact, 
our system features two additional operational modes
that preview the EMS-instructions
using (1) low-intensity EMS (i.e., creates very subtle movements) or (2) sub-threshold EMS (i.e., the simulation creates electro-tactile feedback that suggests movements without physical limb
displacements). 

\section{Implementation}\label{implementation}

To help readers replicate our design, we provide technical details in
this section and open-source all of our codebase in our repository\footnote{Code available at \href{https://embodied-ai.tech}{https://embodied-ai.tech}} (also available in the \emph{Supplementary
Material}). Furthermore, to illustrate key aspects of our implementation,
we provide system diagrams and, in \hyperref[appendix]{\emph{Appendix}}, examples that
showcase the contribution of a particular module/feature (all examples
in this section were created using our complete system; for our
technical evaluation, see \hyperref[technical-evaluation-via-an-ablation-study]{\emph{Ablation Study}}.

\begin{figure}[h]
\centering
 \vspace*{-0.3cm}
  \includegraphics[scale=0.3]{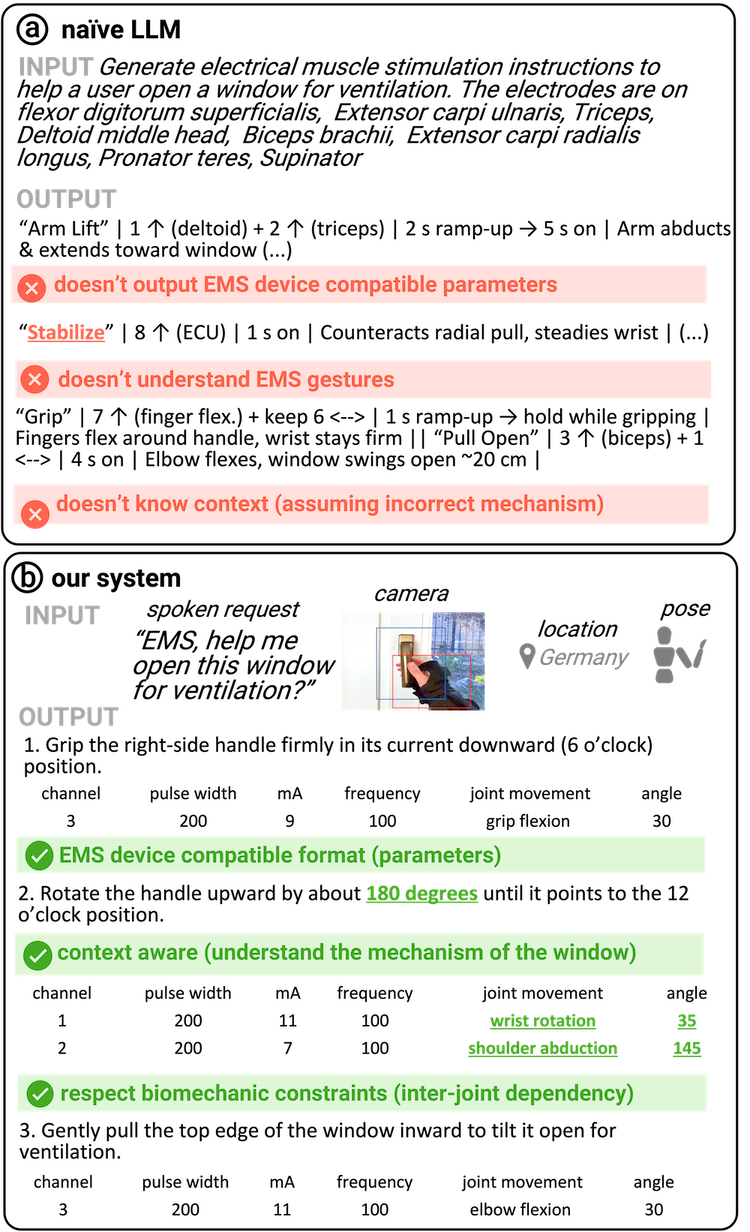}
  \vspace*{-0.3cm}
  \caption{Juxtaposition of naïve OpenAI GPT-4.1 vs. our system.}
  \vspace*{-0.4cm}
  \Description{Two examples of output, left is one from a naïve GPT o1, right is the output of our system. In green and red are highlighted the differences, these follow the same textual description as in the paper. Succinctly, the naïve model did not output EMS compatible parameters, did not understand certain gestures, and assumed the incorrect mechanism for the task (window). Conversely, at the bottom all these three items are shown in green, indicating success.}
  \label{fig:7}
\end{figure}

First, to illustrate the difference between naïvely generating
EMS-instructions using an LLM vs. by leveraging our system,
\hyperref[fig:7]{Figure 7} provides detailed output from the
\hyperref[walkthrough]{\emph{Walkthrough}} (i.e., opening the window to air the room).

\begin{figure}[h]
\centering
  \includegraphics[width=0.95\columnwidth]{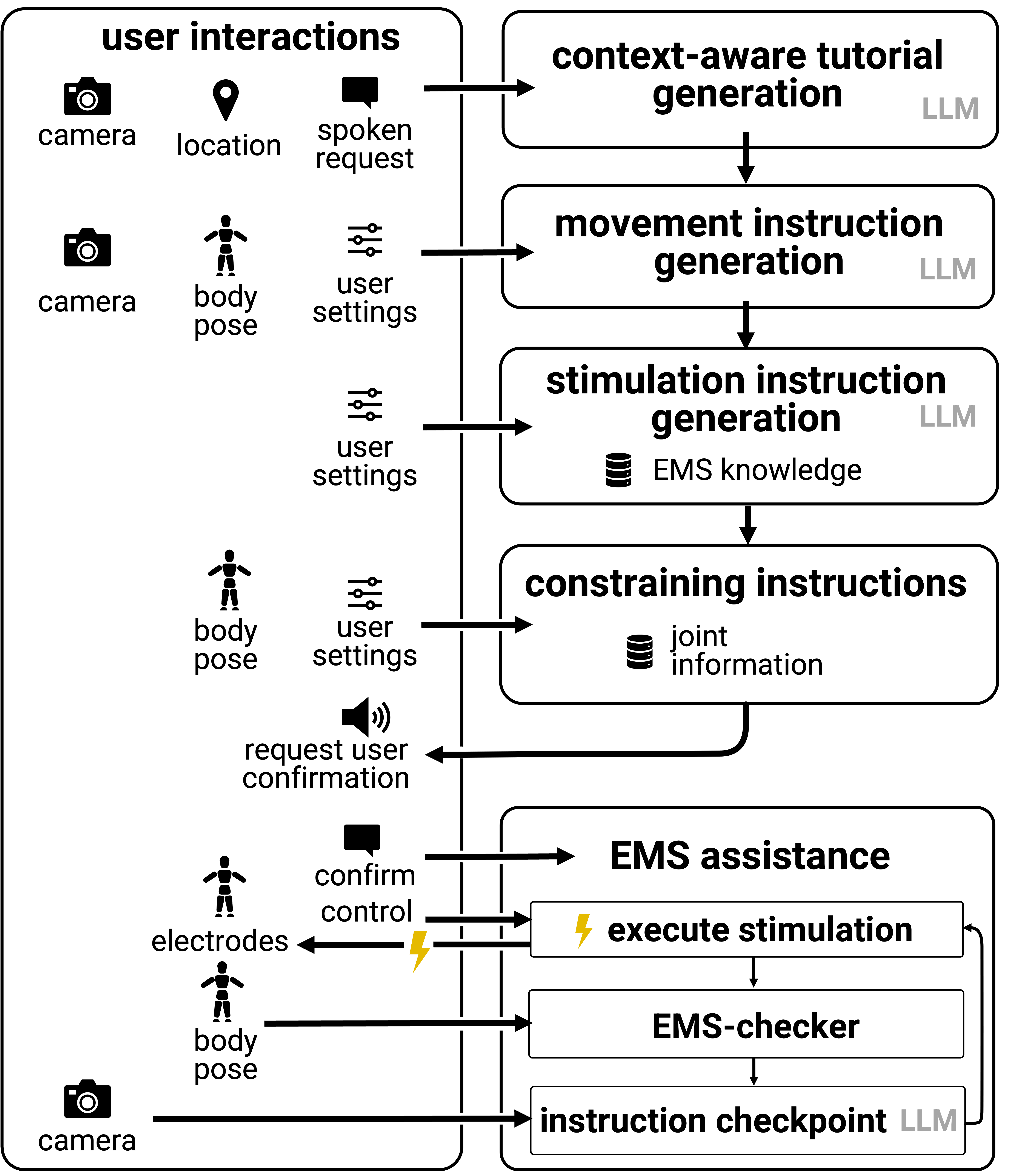}
  \vspace*{-0.3cm}
  \caption{Outline of our system’s architecture and each of its implemented modules.}
  \vspace*{-0.4cm}
  \Description{Outline of our system’s modular architecture. The same five key blocks described in the paper are visible and connected with arrows that suggest information flow from the user’s prompt to EMS-instructions.}
  \label{fig:8}
\end{figure}

In both cases above, the LLM is the same (\emph{OpenAI GPT-4.1})\emph{;}
however\emph{,} the naïve case depicts how foundation models excel at
general-reasoning but fail to consider EMS-knowledge (e.g., incorrect
use of wrist extensor), contextual-cues (e.g., failed to identify the
tilt-turn window), or joint-information (e.g., current wrist pose).
Adding these EMS considerations is a technical challenge that our system
solves. \hyperref[fig:8]{Figure 8} details the system modules
needed to implement these capabilities. Also, prompts and
EMS-knowledge-base are in \emph{Supplementary Material}.

\subsection{User interactions}\label{user-interactions}

\begin{figure}[h]
\centering
    \vspace*{-0.3cm}
  \includegraphics[width=\columnwidth]{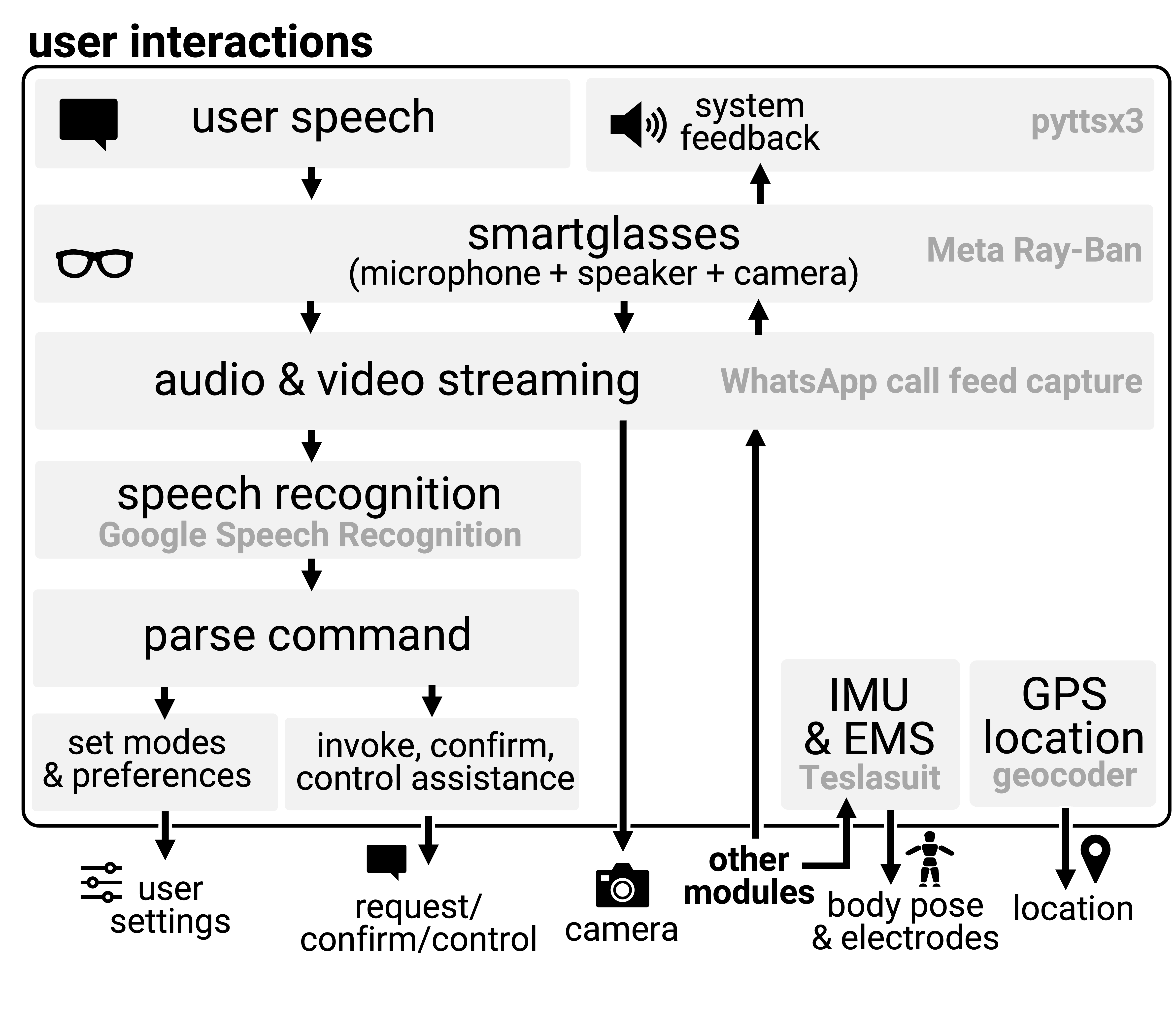}
  \vspace*{-1.2cm}
  \caption{Information flow in the User interactions module.}
  \Description{Flowchart depicting information flow in the user interactions module—image follows textual description in paper.}
  \label{fig:9}
  \vspace*{-0.4cm}
\end{figure}

As depicted in \hyperref[fig:9]{Figure 9}, this module handles
interactions between users and submodules. By piping video and audio
from the smart-glasses (similar to \cite{hillTwoStudentsCreated2024}),
our system transcribes speech-to-text (upon detection of the ``EMS''
wake-up word). Using speech, users can make requests (e.g., ``EMS, help
me with this'') or adjust the delivery of EMS-instructions (e.g., saying
``repeat'', ``slow down/speed up'', ``pause'', ``stop'', ``resume'').
Before EMS takes place, the system reads out a summary of the soon-to-be
stimulated muscles and awaits user confirmation.

\textbf{Advanced settings.} Users can also configure how the stimulation
operates. They can adjust the stimulation mode by saying ``EMS
stimulation mode: ``\emph{actuate}, \emph{nudge}, or \emph{tactile}.''
These modes change the overall intensity of the stimulation: ``actuate''
sets the stimulation at the maximum intensity calibrated by the user,
``nudge'' sets it at a medium-intensity value, and ``tactile'' applies
sub-motor threshold stimulation (not strong enough to cause an
involuntary contraction). Users can also change how the system behaves
when movements are beyond EMS capabilities by saying ``EMS, completion
mode: \emph{partial} or \emph{full}''. In partial-mode, our system will
execute all instructions but skip the ones that it cannot find a
suitable EMS-instruction in our EMS-knowledge base (e.g., if there are
no electrodes on the left hand, but the task would require left-hand
movements)---in this case, it will still say these instructions
out loud. Conversely, in the full-mode, the system will stop if the
sequence contains any instructions without an EMS counterpart. Finally,
users can add custom settings, which are saved in the system's prompt
(e.g., ``EMS, user-settings: my right hand has a weak grasp'').

\subsection{\texorpdfstring{Context-aware tutorial generation
}{Context-aware tutorial generation }}\label{context-aware-tutorial-generation}

\begin{figure}[h]
\centering
  \vspace*{-0.3cm}
  \includegraphics[width=\columnwidth]{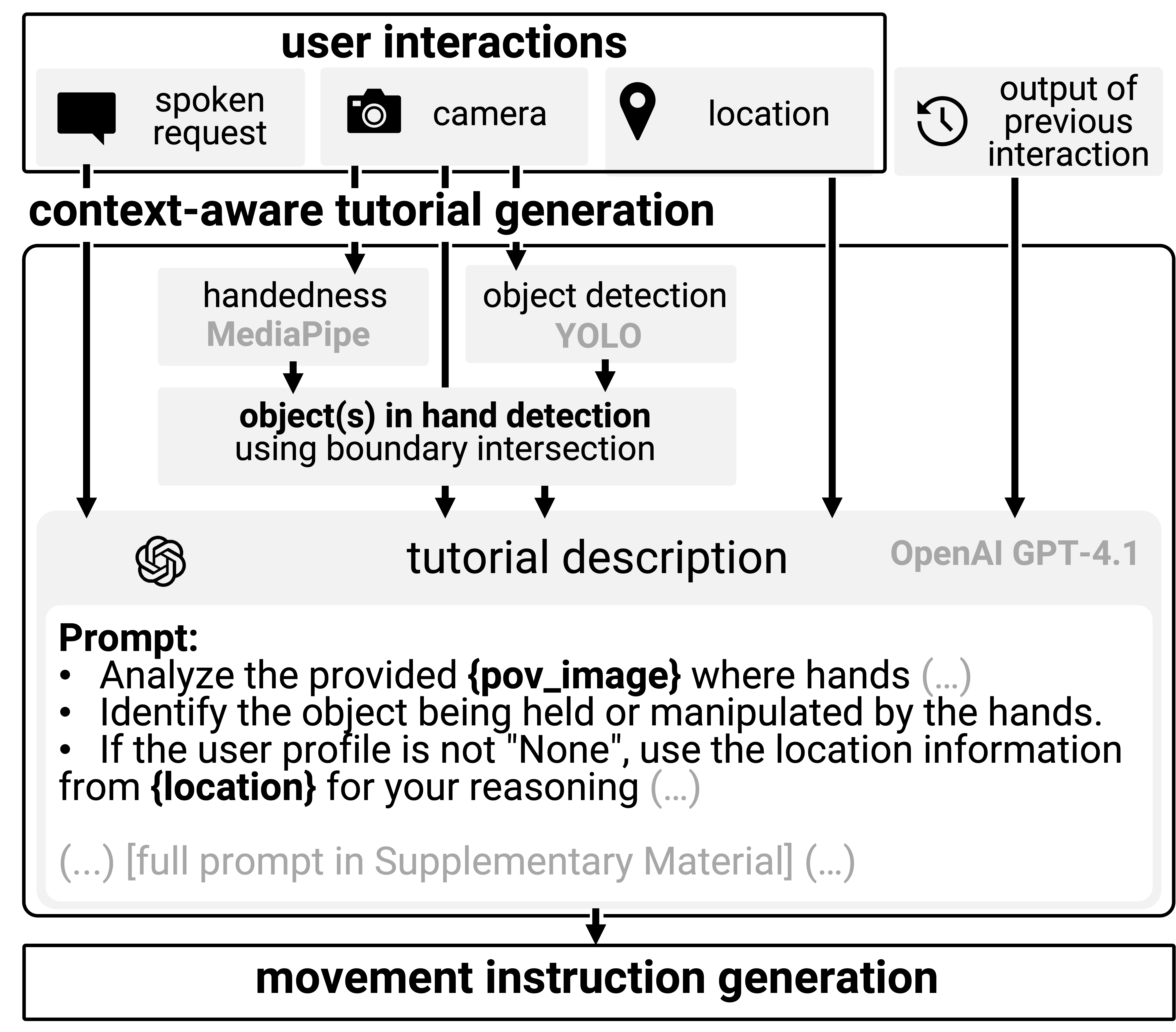}
  \vspace*{-0.7cm}
  \caption{Information flow in the Context-aware module.}
  \vspace*{-0.4cm}
  \Description{Flowchart depicting information flow in the context-aware module—image follows textual description in paper.}
  \label{fig:10}
\end{figure}

\hyperref[fig:10]{Figure 10} depicts the flow of information in
this module, which uses multimodal-AI to \emph{generate} tutorial-like
descriptions to perform the requested task. Rather than just taking the
user's spoken request, it also appends the following (if available): POV
image + location + output of previous interaction. Most of the
contextual-analysis is done by the LLM (OpenAI \emph{GPT-4.1}), but to
infer object-in-hand, we leverage hand tracking (\emph{MediaPipe)} and
object classification (\emph{YOLO}) as LLMs are statistically more
biased \cite{huGenerativeLanguageModels2025} (e.g., only 10.5\% of the
population is left-handed
\cite{papadatou-pastouHumanHandednessMetaanalysis2020} and LLMs could
hallucinate right-handed instructions, even if an object was grasped on
the left). See \hyperref[appendix]{\emph{Appendix}} for comparisons of contextual-analysis
outcomes with and without GPS locations, visual contexts, and user
requests.

\subsection{\texorpdfstring{Movement-instruction generation
}{Movement-instruction generation }}\label{movement-instruction-generation}

\begin{figure}[h]
\centering
 \vspace*{-0.3cm}
  \includegraphics[width=\columnwidth]{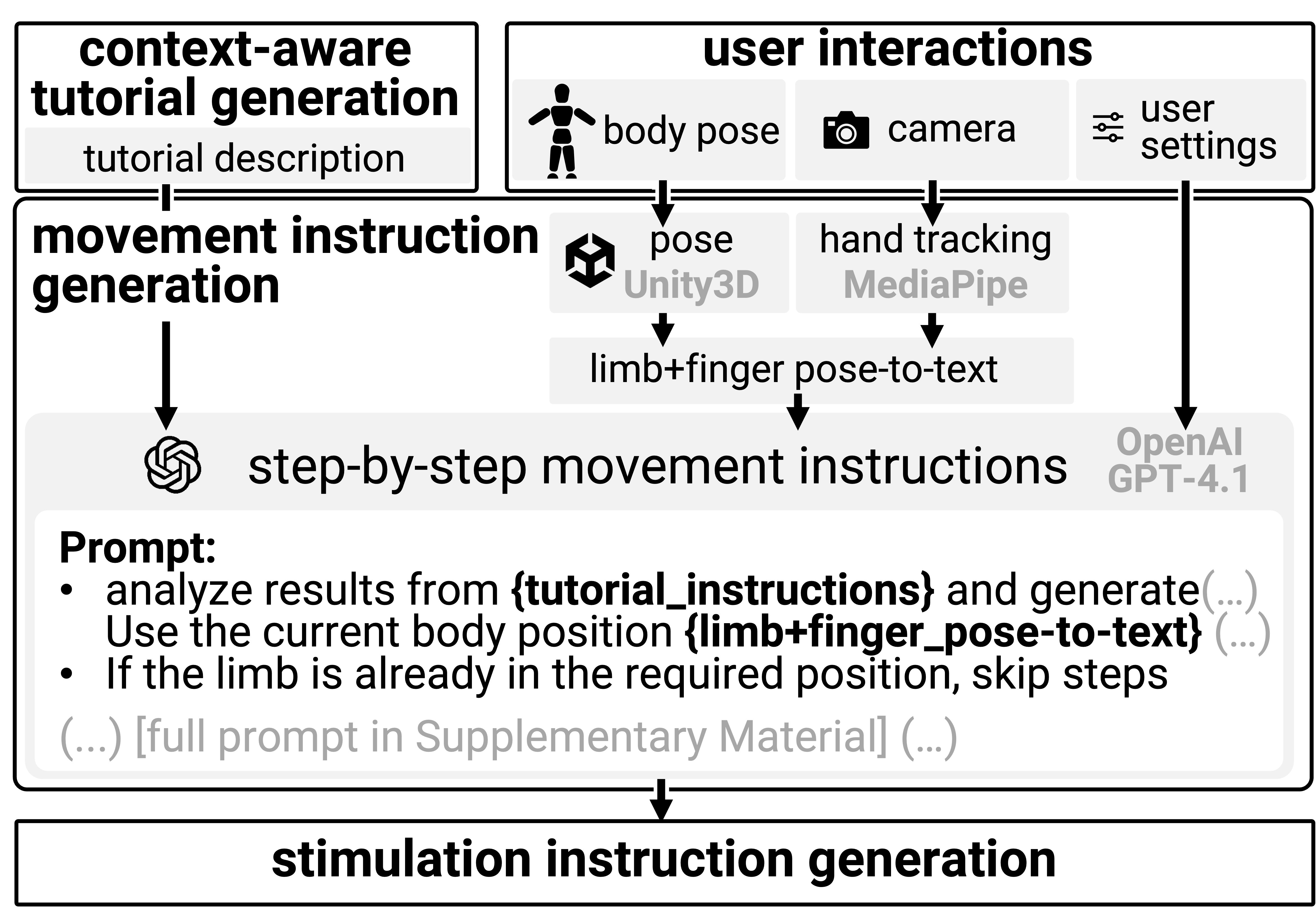}
  \vspace*{-0.5cm}
  \caption{Information flow in Movement instruction generation module.}
  \vspace*{-0.4cm}
  \Description{Flowchart of information flow in the movement module—image follows textual description.}
  \label{fig:11}
\end{figure}

\hyperref[fig:11]{Figure 11} depicts this module, which uses an
LLM to parse the output of the previous module (i.e., a tutorial
description) and generate a sequence of specific movements, taking into
account body-pose and any added preferences. Current body pose is
obtained by a mocap-suit (\emph{Teslasuit}) and by the glasses' camera
(\emph{MediaPipe}) for hand pose. Then, to convert poses to text (for
LLM), we utilize a rule-based system that describes the pose of each
limb along an X-Y-Z axis projected from the user's egocentric-view
(e.g., ``palm up'' if the palm faces the ceiling, and so forth), all
computed in Unity3D. See \hyperref[appendix]{\emph{Appendix}} for a side-by-side example of
generated movement instructions with and without body pose.

\subsection{\texorpdfstring{Stimulation instruction generation
}{Stimulation instruction generation }}\label{stimulation-instruction-generation}

\begin{figure}[h]
\centering
  \includegraphics[width=\columnwidth]{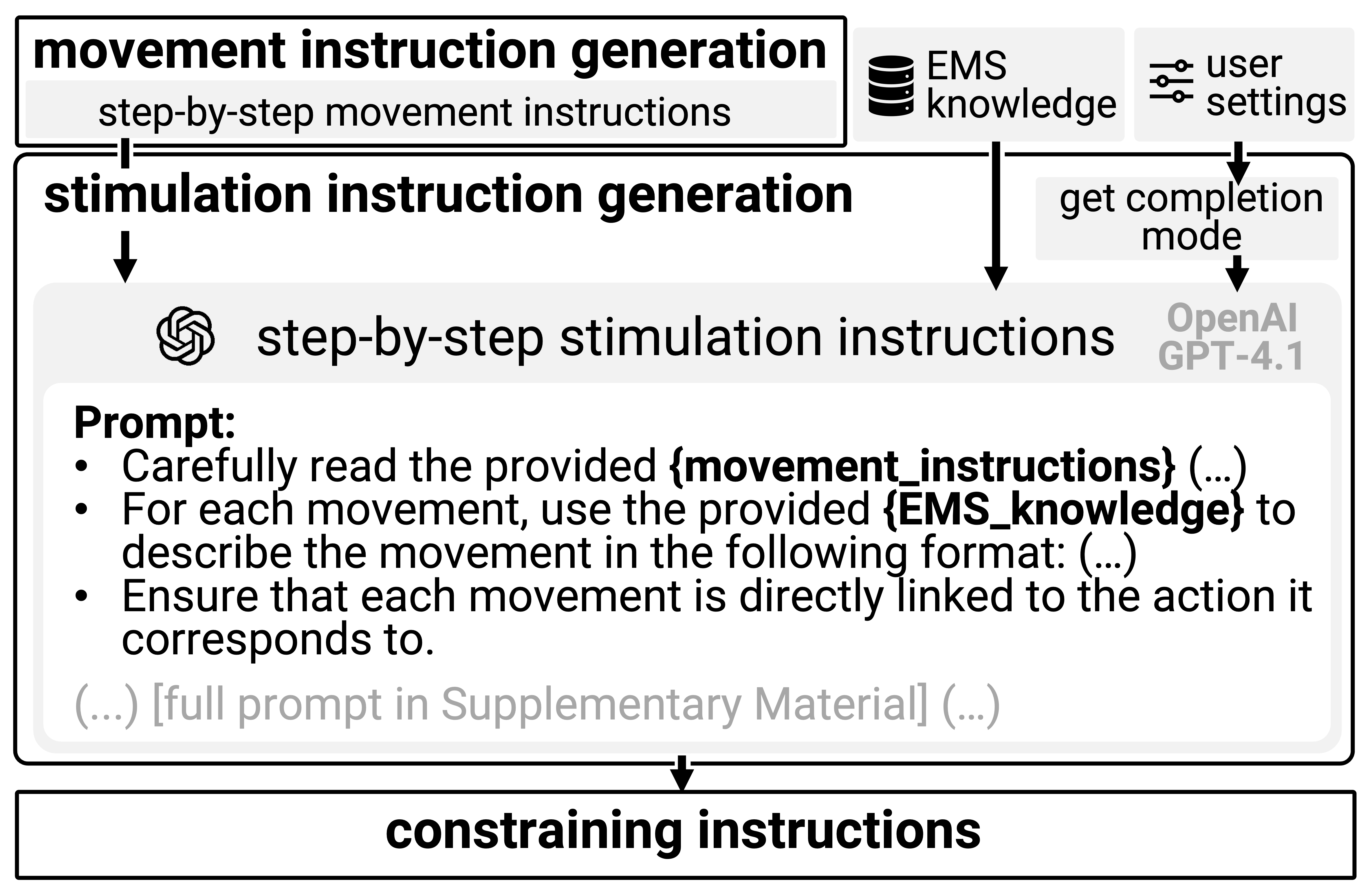}
  \vspace*{-0.5cm}
  \caption{Information flow in Stimulation instruction generation module.}
  \Description{Flowchart depicting the information flow in the EMS-instruction module—image follows text in this case.}
  \label{fig:12}
  \vspace*{-0.4cm}
  
\end{figure}

This module uses an LLM to \emph{select} a set of EMS-instructions from
a knowledge-base of feasible EMS-actuations. As depicted in
\hyperref[fig:12]{Figure 12}, it achieves this by taking the
output of the previous module (i.e., movement-based instructions), and
for each of these, it selects any number of gestures from the
EMS-actuations knowledge-base to achieve this movement instruction. It
is worth noting that there are a number of technical ways to achieve
this step (including without LLMs, using more traditional text-filtering
methods).

Also, as depicted in \hyperref[fig:7]{Figure 7}, without such an
EMS knowledge-base, a naïve-LLM will not output valid EMS commands that
a stimulator can execute (as these have specific EMS-stimulation
parameters, e.g., pulse-width, amplitude, etc.).

\subsection{Constraining instructions with
EMS-knowledge}\label{constraining-instructions-with-ems-knowledge}

\begin{figure}[h]
\centering
  \includegraphics[width=0.95\columnwidth]{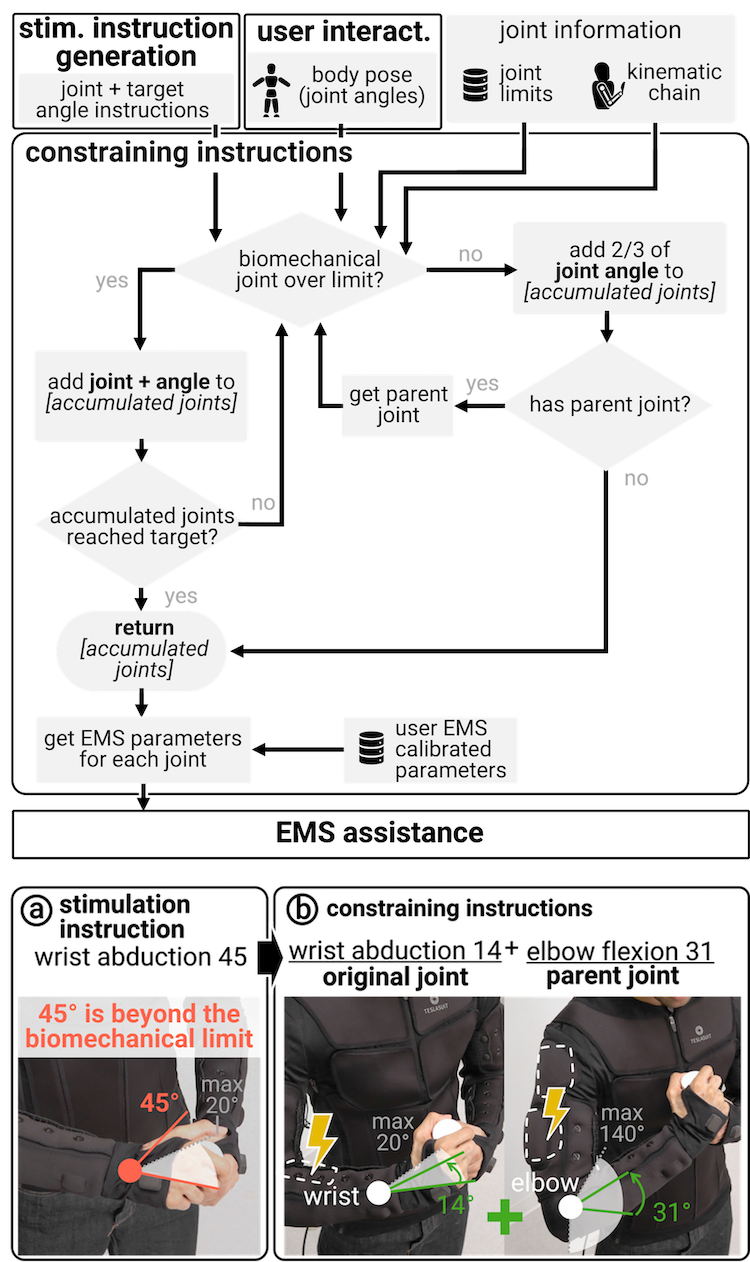}
  \vspace*{-0.3cm}
  \caption{Information flow in the Constraining module and an example of joint constraints in action: (a) without constraints, a generated EMS-instruction might actuate a joint beyond its limits (e.g., wrist past 45°); conversely, (b) the constraining module distributes the angles over the kinematic chain (e.g., moves the wrist by 14° and the elbow by 31°—adding up to the intended 45°).}
  \Description{“(a) Flowchart depicting the information flow in the constraining module—image follows text in this case. (b) Depiction of two users trying to open a pill bottle cap by rotating it. The left user receives EMS on the wrist, but the joint is maxed out. The right user receives EMS on both the wrist and the elbow, allowing them to perform the complete 45-degree movement.”}
  \vspace*{-0.4cm}
  
  \label{fig:13}
\end{figure}

This module uses a \textbf{rule-based system} (no-LLM) to \emph{modify}
the output of the previous module (i.e., sequence of stimulation
instructions), and takes into account hard-constraints (i.e., a table of
joint-limits obtained from
\cite{cazacuStructuralKinematicAspects2014,chenMetaAnalysisNormativeCervical1999,Physiopedia,WristHandActive},
the current pose from our tracking, and a list of the human joints as a
kinematic-chain) as depicted in \hyperref[fig:13]{Figure 13}. For
each stimulation instruction: (1) \emph{stop} the instruction if it is
not biomechanically possible (e.g., stops if the EMS-instruction
requires ``turn neck, clockwise, 180 degrees''); (2) \emph{stop} the
instruction if the joint's current-angle would not allow it (e.g., stops
for the instruction ``turn neck, clockwise, 10 degrees'' if the neck is
already at its maximum angle); and, more generally-speaking, (3)
\textbf{attempts to \emph{adjust} the instruction} if there is a parent
joint (i.e., the joint that is closer to the torso from the current
joint) in the kinematic-chain. If this is the case, it adds up the
angles of successive movements in upward joints until the target angle
was reached or is deemed impossible---for instance, for the instruction
``wrist, abduction, 45 degrees'' while the user's wrist is
fully-extended, it would proceed to the parent joint (i.e., elbow) and
examine its current pose (and if not maxed out) would extend it by 45
degrees. This repeats iteratively up the entire kinematic-chain. As we
show in our ablation study, this adjustment-step is important since a
multimodal-AI might ignore or hallucinate some pose aspects (e.g.,
individual joint-limits, assume poses, etc.) during instruction
generation.

\subsection{EMS assistance}\label{ems-assistance}

\begin{figure}[h]
\centering
\vspace*{-0.3cm}
  \includegraphics[width=\columnwidth]{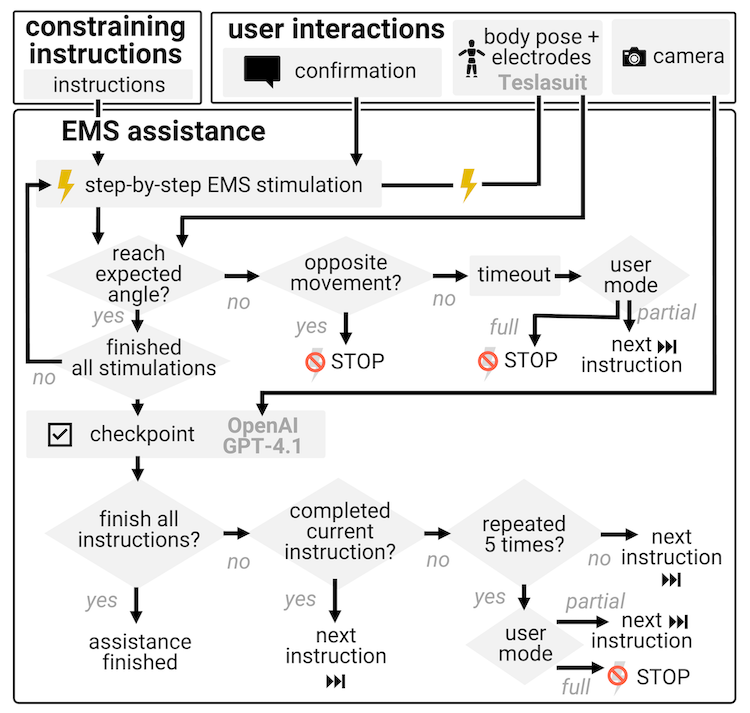}
  \vspace*{-0.8cm}
  \caption{Information flow in the EMS assistance module.}
  \Description{“Flowchart depicting the information flow in the EMS assistance module—image follows text in this case.”}
  \label{fig:14}
\end{figure}

This module, depicted in \hyperref[fig:14]{Figure 14}, uses a
rule-based system and an LLM to, respectively, check if the EMS is
moving a joint, and determine if instruction-checkpoints were reached.

\textbf{Execute stimulation.} EMS is delivered via a \emph{Teslasuit}
with 80 EMS channels: 16 per each of the four limbs and 16 on the torso
(see electrode-placement map in \cite{TeslasuitSelfstandingFES}) which 
is controlled using \emph{Teslasuit}'s \emph{Unity3D} API
\cite{TeslaSuitDeveloper}. Like most EMS devices, this API allows
control of the frequency, amplitude, pulse-width, and duration of a
stimulation---these are per-user pre-calibrated settings for the desired
muscle (see \emph{Constraining} module).

\textbf{EMS-checker.} From the currently actuated-joint and its
target-direction, a rule-based system checks if this joint is:
\textbf{(1) moving in the expected direction:} system continues;
\textbf{(2) not moving significantly} against a pre-calibrated
IMU-threshold: after 3-seconds without movement, EMS is halted and
depending on the user-selected completion mode, this stimulation is
skipped or the entire interaction stops; or, \textbf{(3) moving in the
opposite direction}: user opted to cancel the EMS by moving in the
direction opposite to EMS-actuation, which stops the interaction. This
applies when a collision occurs, as the observed movement is not the
same as the instructed one.

\textbf{Checkpoints.} After all the EMS-instructions for this movement
instruction have been delivered, our system takes a photo of the outcome
and uses its LLM to select the next movement instruction (from the
previously generated list). This allows the system to engage in a
``checkpoint''-like behavior, where it can skip instructions if two steps
were achieved in one-go (e.g., potentially due to user's own intervention), or
even repeat instructions up to five times (e.g., as seen when
the pill bottle cap did not open in the \hyperref[walkthrough]{\emph{Walkthrough}}).

\subsection{Additional
safety-considerations}\label{additional-safety-considerations}

Apart from joint-limit considerations, we implemented typical EMS
safety-considerations (e.g., movements calibrated per-user to operate
pain-free). Further, the \emph{user interactions} module runs in a
separate process, providing always-available control (i.e., ``EMS
pause/stop'' or moving against EMS-actuation, detected by body-pose
tracker). Finally, interactions via our full system are always
user-initiated and require user-confirmation before stimulation.

\vspace*{-0.4cm}
\section{Technical evaluation via an ablation
study}\label{technical-evaluation-via-an-ablation-study}

In this evaluation, we break down the contributions of the modules
through an ablation study, where we compare instructions generated by
one version of our system (e.g., complete system, ablation of a module,
and a Naïve-VLM with our EMS-knowledge base) against a set of
ground-truth EMS-instructions. To accelerate research in this area, we
provide ground-truth and generated EMS-instructions in
\emph{Supplementary Materials}---this is unprecedented in this thriving
research area, as no such datasets exist.

\subsection{Study design}\label{study-design}

\begin{figure}[h]
\centering
  \includegraphics[width=\columnwidth]{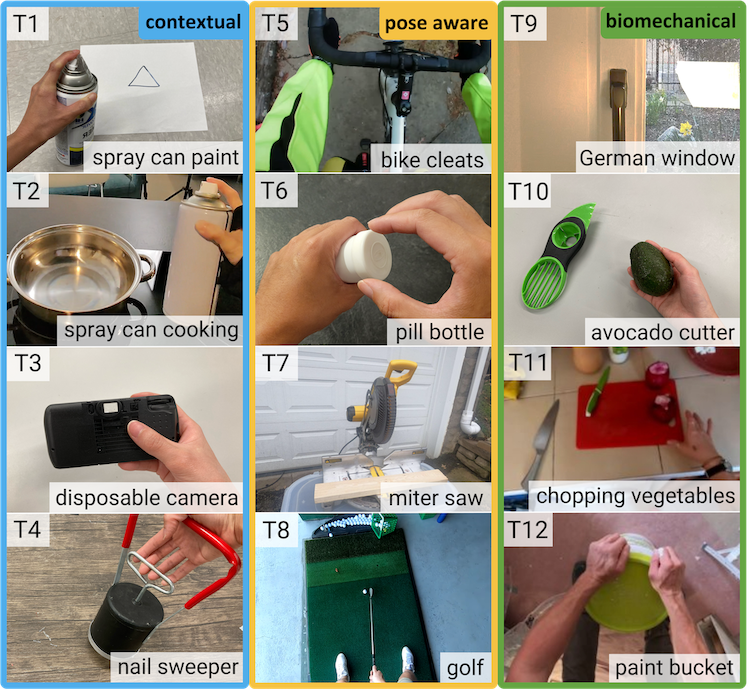}
  \vspace*{-0.6cm}
  \caption{Our 12 physical tasks used in the ablation study, including, (T1) spray-can for painting, (T2) spray-can for cooking oil, (T3) using a disposable camera with film rewinding, (T4) picking up nails with a magnetic-sweeper, (T5) unclipping right foot from a clip-based bike-pedal, (T6) opening push-twist pill bottle, (T7) using a miter-saw, (T8) positioning left hand on golf-club, (T9) open a tilt -turn window (e.g., “German window”), (T10) remove avocado’s pit, (T11) cutting vegetables, and (T12) opening a bucket of paint.}
  \Description{Photos in a grid, each shows a hand interacting with various objects, including spray can, camera, nail sweeper, bike, pill bottle, miter saw, golf, German windows, avocado cutter, chopping vegetables, paint bucket.}
  \label{fig:15}
  \vspace*{-0.4cm}
\end{figure}

\textbf{Ground truth data.} These EMS-instructions were handwritten by
the authors before the evaluation. The three authors authoring these
instructions jointly have 23 years of experience with EMS. For each
task's starting point (a photo of a user's hand poses and the objects
they see in front), we designed the smallest possible set of
EMS-instructions to deliver procedural-knowledge required for the
task---as mentioned, accuracy of EMS is not matched with that of humans
\cite{doucetNeuromuscularElectricalStimulation2012}; thus, our system
focuses on gestures that approximate the procedural-knowledge, but do
not match human dexterity. Because the ground-truth is written in EMS
commands, not in plain language, these were refined \& validated using
EMS code (i.e., we posed our bodies as shown in
\hyperref[fig:15]{Figure 15} and applied EMS to match gestures
that would provide procedural-knowledge to a user). This process took
\textasciitilde20 hours. Succinctly, the final instruction sets,
provided in \emph{Supplementary Material}, are structured as follows: T1
(task 1) with three gestures (index, wrist), T2 with one gesture (index
flexion), T3 with four gestures (index, thumb), T4 with eight gestures
(including shoulder, fingers), T5 with two gestures (ankle
eversion/flexion), T6 with nine gestures (including shoulder, hand), T7
with 17 gestures (including shoulder, elbow, hand), T8 with seven
gestures (including shoulder, hand), T9 with nine gestures (including
shoulder, elbow, hand), T10 with 14 gestures (including shoulder, elbow,
hand), T11 with 17 gestures (including shoulder, hand), and T12 with one
gesture (wrist extension).

\textbf{Tasks}. We ran our study on 12 physical tasks, depicted in
\hyperref[fig:15]{Figure 15}, selected from prior work (e.g., EMS
\cite{faltaousGeniePuttAugmentingHuman2021,kasaharaPreemptiveActionAccelerating2019,lopesAffordanceAllowingObjects2015}
and \emph{POV} dataset \cite{graumanEgo4dWorld30002022}) and designed to
illustrate different aspects of procedural-knowledge: (T1-T4) focused on
contextual-clues (e.g., spraying can for painting, spraying can for
cooking, using a disposable film camera, picking up nails with a
magnetic-sweeper), (T5-T8) focused on body-poses (e.g., unclipping
right-bike pedal, opening push-twist pill bottle, using a miter saw with
body-pose out of sight, positioning left-hand on golf club), and
(T9-T12) focused on biomechanical constraints (e.g., opening a tilt-turn
window, removing the pit from an avocado with left hand, cutting
vegetables with left hand, opening a bucket of paint). Four tasks were
adopted from \emph{Affordance++}
\cite{lopesAffordanceAllowingObjects2015} (e.g., tilt-turn window,
avocado tool, magnetic-sweeper, and paint spray-can), four tasks were
from the \emph{ego4D} dataset \cite{graumanEgo4dWorld30002022} (e.g.,
biking, miter saw, cutting vegetables, and paint bucket), three inspired
by prior EMS work (e.g., camera
\cite{kasaharaPreemptiveActionAccelerating2019}, golf
\cite{faltaousGeniePuttAugmentingHuman2021}, and a variation on the
spray-can \cite{lopesAffordanceAllowingObjects2015}), and one crafted by
the authors (e.g., opening a pill bottle). Images required as visual
prompts for all these tasks were taken either directly from the papers
referencing these tasks, including the golf task (T5, T7, T8, T11, and
T12 from \emph{ego4D} dataset \cite{graumanEgo4dWorld30002022}), or
reconstructed by posing the objects on a table alongside the author's
hand pose from a POV---see \hyperref[fig:15]{Figure 15}. User
poses were also reconstructed for each scenario by posing our system's
skeletal-model inside \emph{Unity3D}.

\textbf{Ablation conditions.} Per task, five sets of EMS-instructions
were generated, one per condition: (1) \textbf{full system} (all
modules, except checkpoints, since we only fed the starting position for
each task in \hyperref[fig:15]{Figure 15}); three different
ablated-conditions corresponding to removal of a system's module: (2)
\textbf{contextual ablation} (i.e., removed POV image for reasoning,
object recognition, GPS data, and user instruction were replaced by generic
``help me''), (3) \textbf{user-pose ablation} (i.e., removed user-pose),
and (4) \textbf{EMS-knowledge ablation} (i.e., database contained
EMS-movements but no joint-limits, kinematic-chain, or user preferences
such as dominant-hand); finally, for contrast, we also generated
instructions using a (5) \textbf{naïve-VLM}, our baseline (\emph{OpenAI
GPT-4.1} with user request ``help me''). Outputs were standardized to
\emph{{[}handedness{]}{[}limb{]}{[}movement{]}{[}target angle{]}}. Thus,
even in the naïve-VLM, we added instruction-format and a reduced-version
of our EMS knowledge-base (i.e., only had access to valid EMS movements,
but no joint-limits, etc.); otherwise, the VLM would output text that
could not be understood by an EMS stimulator, nor compared to the
remaining conditions in this study. All conditions had the same inputs,
including contextual (POV image, user's spoken instructions, object
recognition, geographical data), user-pose data (\emph{Unity3D}
generated poses, hand tracking), and EMS-knowledge (joint-limits,
kinematic-chain, possible EMS movements), unless ignored in their
respective ablated-conditions.

\textbf{Latency measurement.} On average, it took our system an average
of 23.6s (SD=6.3) to generate instructions across these tasks, because
it uses three LLM-API calls, compared to the average 7.2s (SD=2.7) of
the Naïve-VLM (one API call).

\subsection{Metrics}\label{metrics}

To compare the differences between generated-instructions vs.
ground-truth, we used a modified \emph{Levenshtein distance}
\cite{levenshteinBinaryCodesCapable1966}. This depicts the
\textbf{minimum cost} for transforming a set of generated instructions
into those of the ground-truth set, accounting for \emph{insertions},
\emph{deletions}, and \emph{substitutions} at the instruction-level.
Each operation was assigned a penalty cost proportional to its impact on
movement fidelity. \emph{Insertions} were assigned a cost of 1, since
additional instructions may embellish or reinforce a motion without
altering it. \emph{Deletions} were assigned a higher cost of 6, as
omitting an instruction is likely to result in an incomplete outcome. As
for \emph{substitutions} (i.e., partially-matched instructions), a
finer-grained comparison was applied by evaluating individual components
of the instruction. We consider the following three metrics, each
assigned weights to reflect their relative importance in the muscle
stimulation output and are cumulative to express the substitution cost:
(1) \textbf{incorrect-handedness} (cost 2) if the generated movement's
instructions occur on the opposite side of the body compared to
ground-truth (e.g., left vs. right elbow); (2)
\textbf{incorrect}-\textbf{joint} (cost 2): if the generated movement's
instructions occur on another joint of the body compared to ground-truth
(e.g., elbow vs. wrist); and, (3) \textbf{incorrect-DOF}: (cost 1): if
the generated movement's instructions actuate the wrong
degree-of-freedom (DOF) for a target muscle (e.g., flexion vs. extension
of the wrist). On top of these operations, two additional
violation-types were added: \textbf{biomechanical violation} (cost 2) if
a movement's angle exceeded the joint-limit, and a \textbf{formatting
violation} (cost 2) when deviating from the required format (e.g.,
specifying an angle range instead of a single value or omitting the
target limb)---malformed instructions were still reformatted when
possible for the sake of the comparison analysis but with the violation
penalty.

Also, since the number and order of instructions may vary from the
generated to the ground-truth, we used dynamic programming to align
neighboring instructions and compute the overall minimal transformation
cost per comparison.

Finally, while we applied weights to reflect the severity of mistakes,
we also report the unweighted distances.

\subsection{Results}\label{results}

\begin{figure}[h]
\centering
\vspace*{-0.1cm}
  \includegraphics[width=0.85\columnwidth]{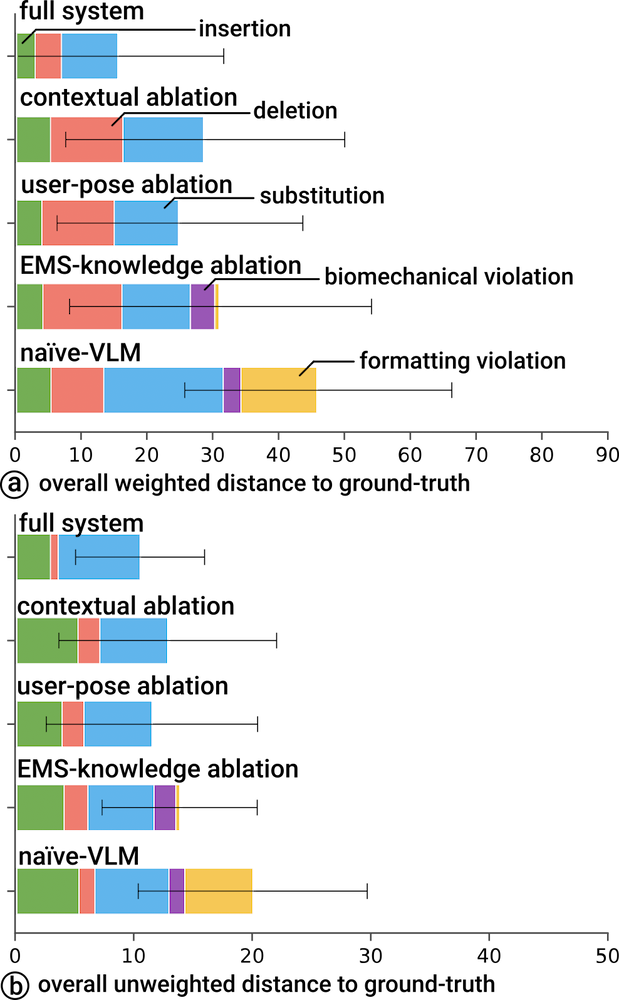}\hfill
  \vspace*{-0.3cm}
  \caption{Our ablation evaluation results for all 12 scenarios.}
  \Description{Two plots comparing weighted and unweighted distance to ground truth from different methods on all 12 scenarios.}
  \label{fig:16}
\end{figure}

\hyperref[fig:16]{Figure 16} depicts our key result. For all 12 tasks, we found that the distance
between \textbf{full system} and \textbf{ground-truth} was the lowest. In
the following, we present the breakdown for all conditions (i.e.,
full system, three ablations, and naïve-VLM) and detail the type
mismatches between generated instructions and ground-truth.

\textbf{Overall.} Our analysis (\hyperref[fig:16]{Figure 16})
revealed the following distances, in descending order (i.e., best to
worst), between generated-instructions vs. ground-truth:
\textbf{full system} with an average of 15.4 (10 unweighted),
\textbf{user-pose ablation} with an average of 24.6 (11 unweighted),
\textbf{contextual ablation} with an average of 28.4 (12 unweighted),
\textbf{biomechanical ablation} with an average of 30.8 (13 unweighted),
and \textbf{naïve-VLM condition} with an average of 45.6 (19
unweighted). These suggest that the full system outperforms the
ablations, and the Naïve-VLM (even when it was assisted with the correct
syntax). Additionally, ablated conditions also outperformed the
Naïve-VLM, showcasing each module's value.

\begin{figure}[h]
\centering
  \includegraphics[width=0.85\columnwidth]{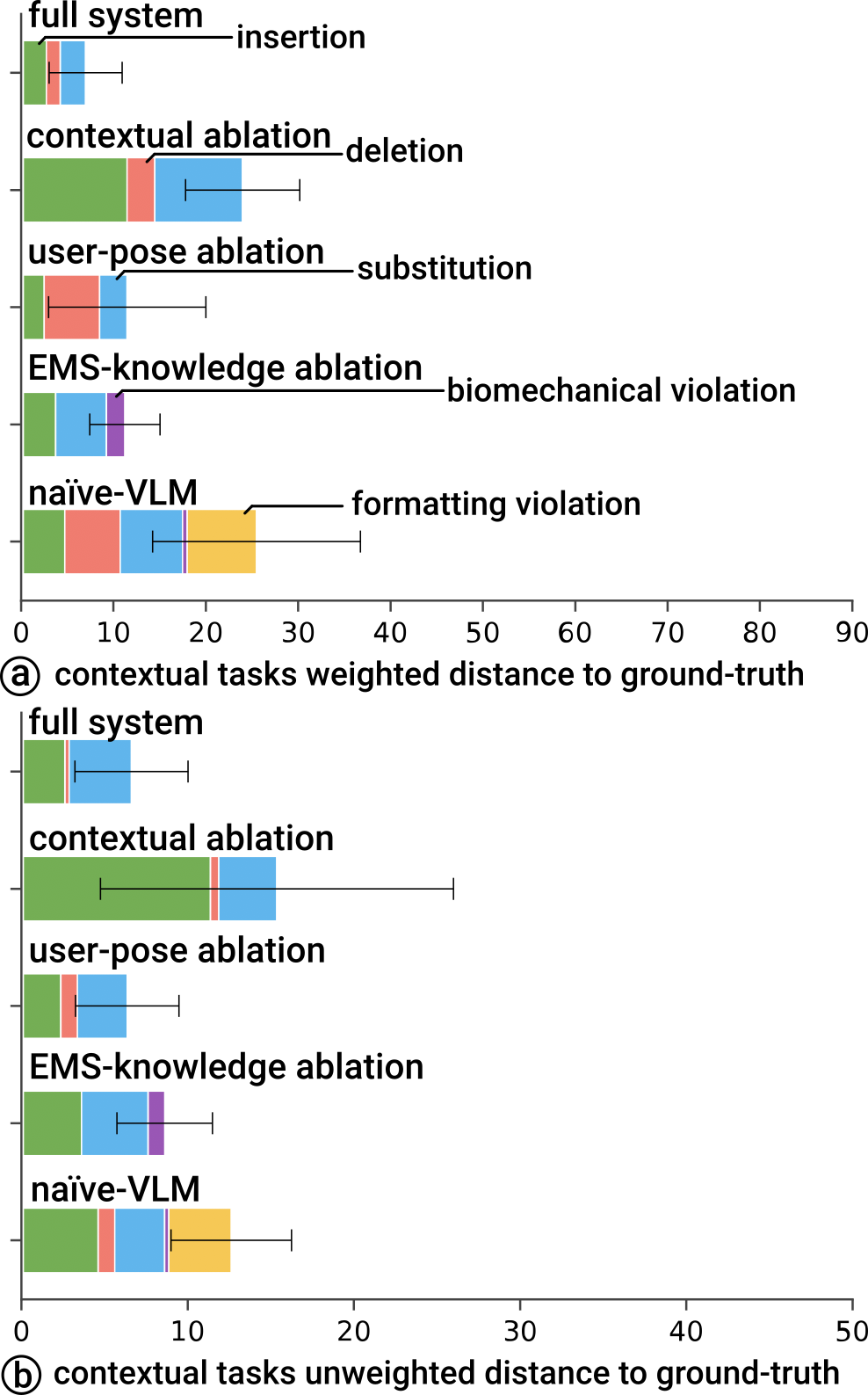}\hfill
  \vspace*{-0.3cm}
  \caption{Our ablation results for all four contextual tasks.\hfill}
  \Description{Two plots comparing weighted and unweighted distance to ground truth from different methods on four contextually relevant scenarios.}
  \label{fig:17}
\end{figure}

\textbf{Contextual-tasks.} In these, we selected scenarios where cues
were useful to derive ground-truth movements.
\hyperref[fig:17]{Figure 17} reveals the following distances, in
descending order: \textbf{full system} with an average of 6.8 (6
unweighted), \textbf{EMS-knowledge ablation} with 11.0 (8 unweighted),
\textbf{user-pose ablation} with 11.3 (6 unweighted), \textbf{contextual
ablation} with 23.8 (15 unweighted), and \textbf{naïve-VLM} with 25.3
(12 unweighted). While performing slightly better than the naïve-VLM,
the contextual ablation performed the worst in terms of raw operation
count, mostly due to \emph{inserting} more instructions than
needed---this was expected, since contextual cues were removed in that
condition. To illustrate one such exemplary task, we focus on the
disposable-camera task (T3), where a user takes a photo but must first
advance the film. In this case, when the contextual-module was ablated
(no POV or object recognition), the system failed to recognize that the
camera was a disposable analog model requiring film advancement.
Instead, it assumed just pressing the shutter. By contrast, all other
conditions correctly instructed the user to advance the film before
taking a photo.

\begin{figure}[h]
\centering
\vspace*{-0.1cm}
  \includegraphics[width=0.9\columnwidth]{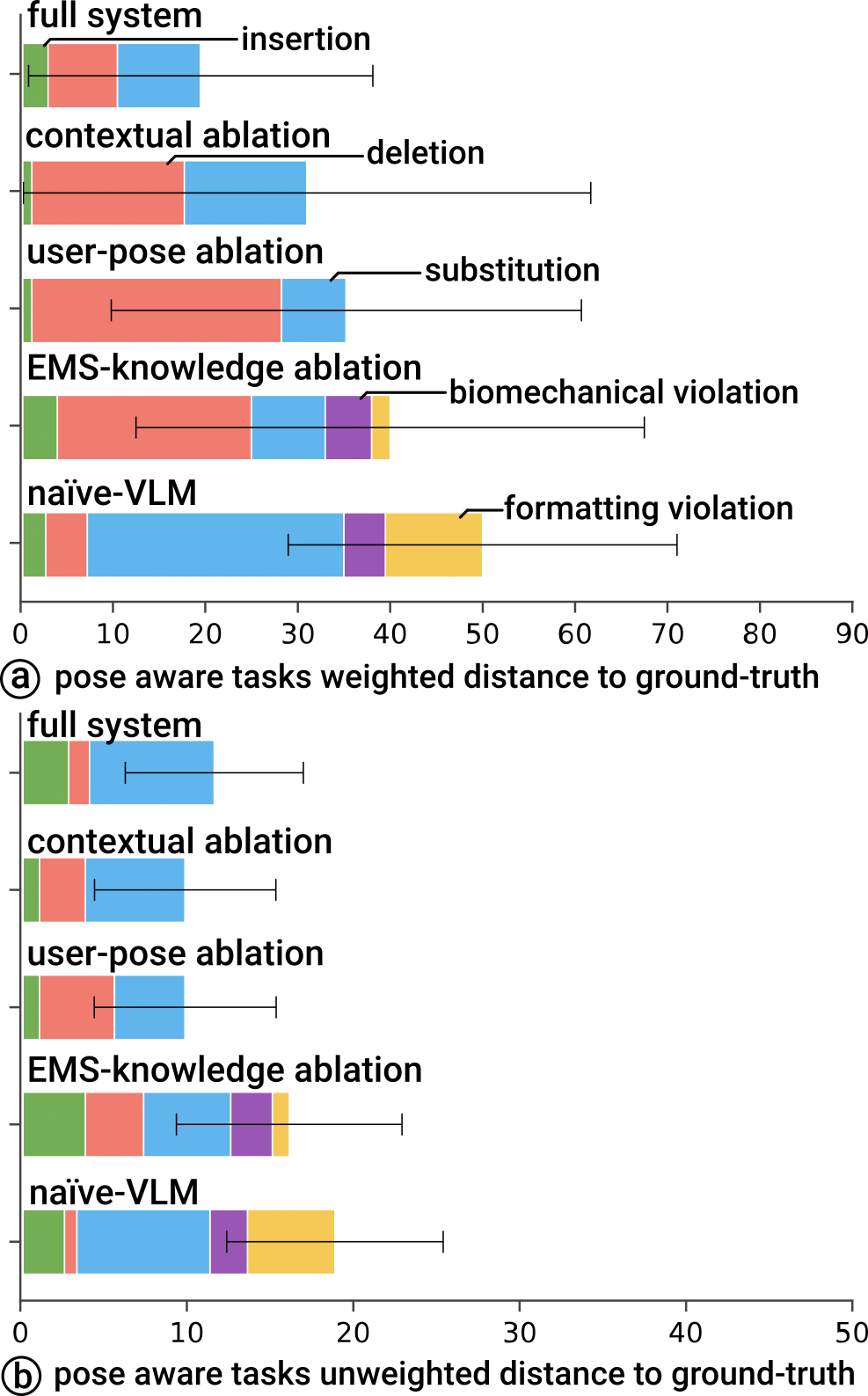}\hfill
  \vspace*{-0.2cm}
  \caption{Our ablation results for all four pose-aware tasks.}
  \vspace*{-0.2cm}
  \Description{Two plots comparing weighted and unweighted distance to ground truth from different methods on four user-pose relevant scenarios.}
  \label{fig:18}
\end{figure}

\textbf{Pose-aware-tasks.} In these, we examined scenarios where access
to pose information was useful for generating correct instructions.
\hyperref[fig:18]{Figure 18} depicts the average edit distance,
in descending order: \textbf{full system} with an average of 19.3 (11
unweighted), \textbf{contextual ablation} with 30.8 (9 unweighted),
\textbf{user-pose ablation} with 35.0 (9 unweighted),
\textbf{EMS-knowledge ablation} with 39.8 (16 unweighted), and
\textbf{naïve-VLM} with 49.8 (18 unweighted). While the EMS-knowledge
ablation had the highest distance amongst the ablated models, a portion
of the penalty was attributed to biomechanical and formatting
violations; without those, the user-pose ablation had the highest
score---this is mostly due to the \emph{deletion} operations (missing
instructions), which could be attributed to the lack of body-pose
knowledge. To illustrate one such exemplary task, we take a closer look
at T8 (i.e., place left-hand on a golf club, as the right-hand is
already on the club). In the user-pose ablation condition, where no pose
information was available to the system, it assumed that both hands were
already on the club---even when provided with a POV image---and
therefore produced no instruction (no correction was deemed necessary).
By contrast, all other systems (except for naïve-VLM) suggested
actuating the left-hand.

\begin{figure}[h]
\centering
\vspace*{-0.3cm}
    \includegraphics[width=0.9\columnwidth]{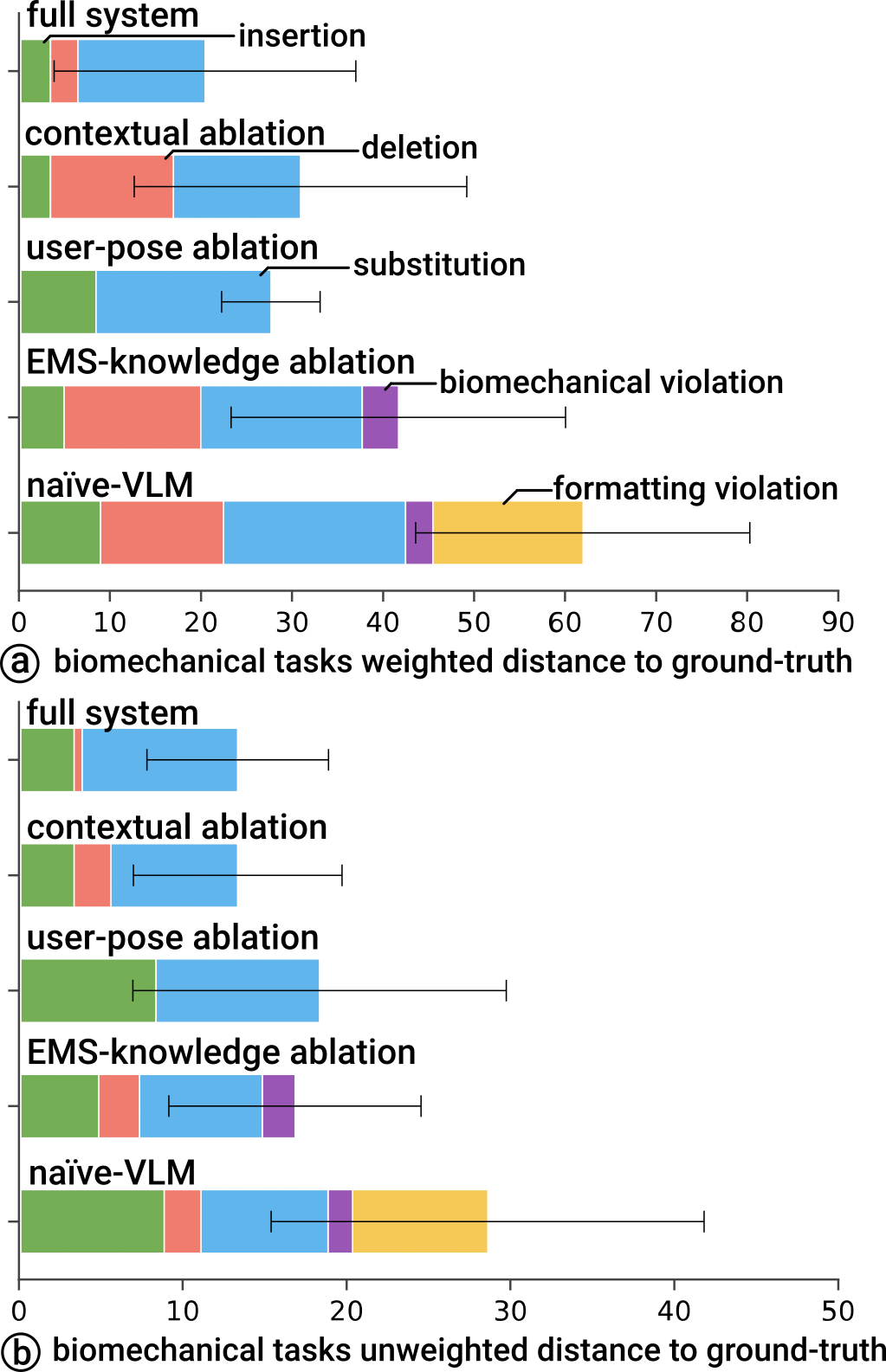}\hfill
    \vspace*{-0.2cm}
    \caption{Our ablation evaluation for all four biomechanical tasks.}
    \vspace*{-0.2cm}
    \Description{Two plots comparing weighted and unweighted distance to ground truth from different methods on four biomechanically relevant scenarios.}
  \label{fig:19}
\end{figure}

\textbf{EMS-knowledge-tasks.} We examined scenarios where EMS knowledge
was useful for generating correct movements.
\hyperref[fig:19]{Figure 19} revealed the following distances, in
descending order: \textbf{full system} with an average of 20.3 (13
unweighted), \textbf{user-pose-ablation} with 27.5 (18 unweighted),
\textbf{contextual ablation} with 30.8 (13 unweighted),
\textbf{EMS-knowledge ablation} with 41.3 (16 unweighted), and
\textbf{naïve-VLM} with 61.8 (28 unweighted). As expected, only
EMS-knowledge-ablation showed the highest overall distance, largely
driven by biomechanical violations, resulting in the worst among the
ablated-conditions. We further noted that only EMS-knowledge-ablation
and naïve-VLM had biomechanical violations since they were the only
conditions without this knowledge. To illustrate one such exemplary
task, we consider T9 (opening a tilt-turn window). This type of window
can be opened either from the side---by rotating the handle to the right
(3 o'clock position)---or from the top, by rotating the handle upwards
(12 o'clock). In this scenario, the user intends to open the window from
the top, requiring the handle to rotate from 6 o'clock to 12 o'clock
(180° rotation). After correctly identifying this requirement, our
system's LLM suggested rotating the wrist 180°. However, it is
biomechanically impossible to perform this motion using only the wrist.
As expected, the \textbf{EMS-knowledge ablated} condition failed to
catch this joint-limit violation and proceeded with the instruction. By
contrast, all other conditions (except naïve-VLM,) detected this and
compensated (via our constraints implementation) by incorporating
shoulder and elbow movements.

\textbf{Evaluation limitations}. In this evaluation, we measured output
using a distance metric. However, the human body is complex, and
different motion trajectories or joint configurations can still achieve
the same end goal. There are also more possible ways of defining
the ground truth, which is itself subject to the complexity discussed
above. Finally, as is typical with these evaluations, the real-time
aspects of our system were disabled, such as our \emph{checkpoints}
feature.

\textbf{Summary of findings.} We found that our system outperformed the
baseline. Moreover, ablations suggest that each module provides a
net-positive contribution: (1) absence of contextual-cues led to
movement-errors, (2) absence of pose-information generated movements
that did not respect the current body-pose; and, finally, (3) absence of
EMS-knowledge leads to incorrect EMS-instructions, sometimes even
violating physical human constraints (i.e., unergonomic). These
technical results highlight that our system never perfectly matched the
EMS-expert's ground-truth. In fact, next, we will use these types of
errors found in our ablation study, to (purposely) simulate errors in
our user study.

\vspace*{-0.4cm}
\section{User Study}\label{user-study}

In our user study, we examined participants' responses to our system in
a laboratory-setting. Since no interactive system will ever perfectly
capture the intent of all possible users, and certainly any system based
on LLMs is susceptible to hallucinations, we designed this study to
\textbf{examine not only what happens when our system generates the
correct instructions}, but especially, \textbf{what happens when the
generated instructions are \underline{erroneous }}e.g., by injecting
simulated errors, similar to those observed in our technical evaluation,
see \hyperref[technical-evaluation-via-an-ablation-study]{\emph{Ablation Study}}).

To control so that all participants experience the same set of erroneous
instructions, we systematically created different failures-modes by
injecting incorrect instructions (e.g., wrong order, wrong limb,
nonsensical movement, etc.). This allowed us to observe if participants
can not only use our system when its output matches the task, but also in extreme cases where a user is required to be critical of the
muscle stimulation output. This allowed us to ask:

(1) \emph{Are users able to detect errors?}

(2) \emph{Upon detecting errors, are users able to ignore errors to make
sense of partially-correct instructions}

(2) \emph{Upon detecting errors, do users attempt to refine the output
by re-prompting the system?}

(4) \emph{When exploring the system's output, what strategies do users
employ?} (e.g., repeat gestures, mimic, slow down)

\begin{figure*}[b]
\centering
 \vspace*{-0.1cm}
  \includegraphics[width=\textwidth, height = 3.8 cm]{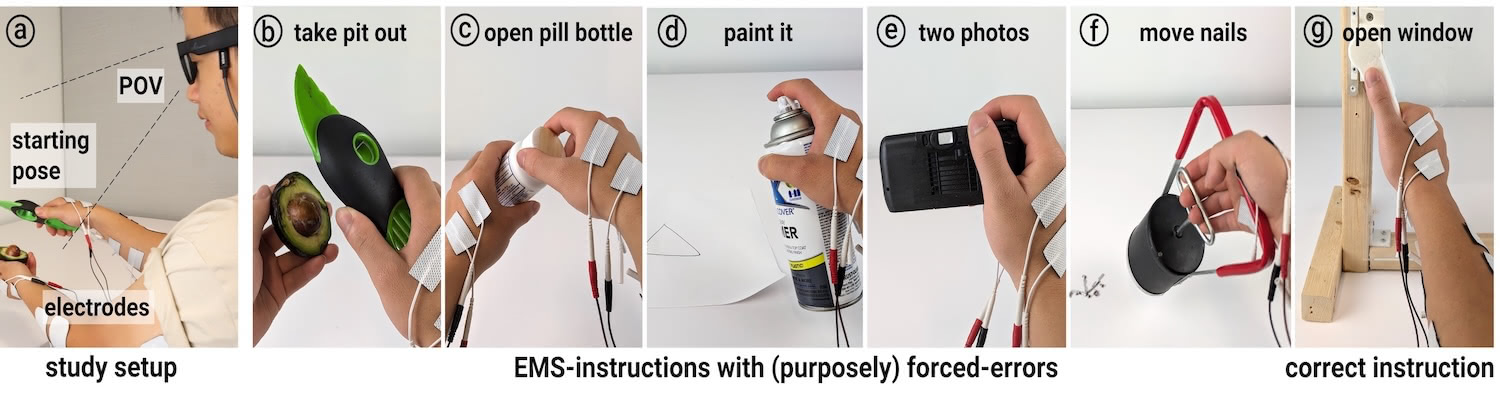}
  \vspace*{-0.7cm}
  \caption{Participants completed six tasks in a random order. (a) shows the study setup. Participants wore smart-glasses and pre-calibrated electrodes and were instructed to start each task with specific starting poses. (b)-(g) shows the starting poses of each task. Task (b)-(f) presented EMS-instructions with purposely forced-errors, and task (g) presented correct EMS-instruction.}
  \vspace*{-0.5cm}
  \Description{Photos show a person wearing glasses and their hands attached with electrodes. A series of photos shows different interactions of the hands with objects.}
  \label{fig:20}
\end{figure*}

\subsection{Study design}\label{study-design-1}

We closely followed the study design of \emph{Affordance++}
\cite{lopesAffordanceAllowingObjects2015}, which was set up to observe
users' reactions to EMS-based physical assistance. In this design, for
each trial, participants are shown objects and asked to perform a task
(e.g., ``paint with a spray-can'') using the assistance of EMS.
Important to this design is that participants are instructed to
think-out-loud---this canonical method is often employed to generate
more insights that are invisible to experimenters, namely to understand
a user's mental model of an
interface\cite{chartersUseThinkaloudMethods2003,hewettUseThinkingoutloudProtocol1987,olsonThinkingOutLoudMethodStudying1984,solomonThinkAloudMethod1995}.

\textbf{User's voice control.} While our complete system allows users to
control it using real-time voice recognition (i.e., saying the wake-up
word ``EMS'' followed by an instruction or command such as ``help me''),
we did not want errors from the voice recognition to interfere with this
study---our goal was to observe participants' reactions to
EMS-instructions, not to measure the accuracy of voice recognition.
Thus, we employed the canonical Wizard-of-Oz
methodology---\emph{Affordance++} also used this
\cite{lopesAffordanceAllowingObjects2015}. The only role the
experimenter denoted as Wizard took was to trigger pre-cached outputs
(i.e., sequences of muscle gestures) as a response to any valid-verbal
commands provided by a participant (e.g., ``EMS, help me'', ``EMS,
repeat'', etc.---as in \hyperref[implementation]{\emph{Implementation Study}}). For instance, when a
participant said ``EMS, help me'', the Wizard initiated the pre-cached
stimulation obtained from our system to that prompt. Similarly, when a
participant said ``EMS, slow down'', the Wizard simply pressed a key
that calls the slow down function in our code. Additionally, since
instructions were pre-cached, this minimized the waiting time of our
system when running online (\textasciitilde23.6s).

\textbf{Instruction generation}. All EMS-instructions were generated by
our complete system ahead of time and pre-cached for each task. To
generate the instructions, our system heard the same verbal prompts
related to the task goals that the participants were instructed to
complete. Alongside this, our system was shown an image of the object in
that trial, including the hands/body-pose of an experimenter next to the
object (and if needed, a user's simulated-location, e.g., Germany, when
the user encountered the tilt-turn window). Participants were asked to
take the starting poses in each task right before the start of trials,
as shown in \hyperref[fig:20]{Figure 20}.

\textbf{Trial.} For each trial, we asked participants to
\emph{think-out-loud} while letting the EMS-instructions guide them,
with procedural knowledge, to perform a task in front of them. At the
end, they rate how well they understand the steps on a 7-item Likert
scale (scale used also in \cite{lopesAffordanceAllowingObjects2015}).

\textbf{Tasks.} \hyperref[fig:20]{Figure 20} (a) depicts our six
tasks, which are simplified versions of tasks from the \hyperref[technical-evaluation-via-an-ablation-study]{\emph{Ablation Study}}: (b) ``take the pit out of the avocado'' (avocado tool)---this
was found to be the most confusing step for the participants in
\emph{Affordance++} \cite{lopesAffordanceAllowingObjects2015}, (c)
``open a pill bottle'' (child-lock pill bottle), (d) ``paint this''
(spray-can), (e) ``take two pictures'' (disposable camera), (f) ``move
the nails to the right'' (magnetic-sweeper), and (g) ``open the window''
(tilt-turn window).

\textbf{EMS-instructions with \underline{forced-errors}.} Important to our
study was to not only observe participants' reactions when our system
generates correct instructions, but especially, when the generated
instructions are not suitable for the task. Thus, we systematically
added five simulated failure-modes that include the errors similar to
those we found in our \hyperref[technical-evaluation-via-an-ablation-study]{\emph{Ablation Study}} and system glitches (e.g.,
tracking-errors): \textbf{(1) Misunderstood-mechanism} (e.g., a
child-lock pill bottle was recognized as a standard pill bottle, thus
the system did not push the cap before turning it)---this simulates an
error in the context-aware tutorial generation module; \textbf{(2)
Wrong-order} (e.g., system pulled the lever of the magnetic-sweeper,
which is a release mechanism, before catching the nails), simulating
movement instruction generation errors. \textbf{(3) Nonsensical} (e.g.,
adding unrelated movements to a task)---simulating hallucinations during
instruction generation; \textbf{(4) Wrong-joint} (e.g., rotation of the
hand with the avocado cutter becomes a rotation of the
neck)---simulating possible hallucinations/tracking-errors; and
\textbf{(5) Wrong-handedness} (e.g., actuating the left hand with the
shaking movements while the participant holds the spray-can in the
right)---simulating hallucinations/tracking-errors.

\vspace*{-0.1cm}
\subsection{Apparatus}\label{apparatus}

\begin{figure*}[b]
\centering
  \includegraphics[scale = 0.1]{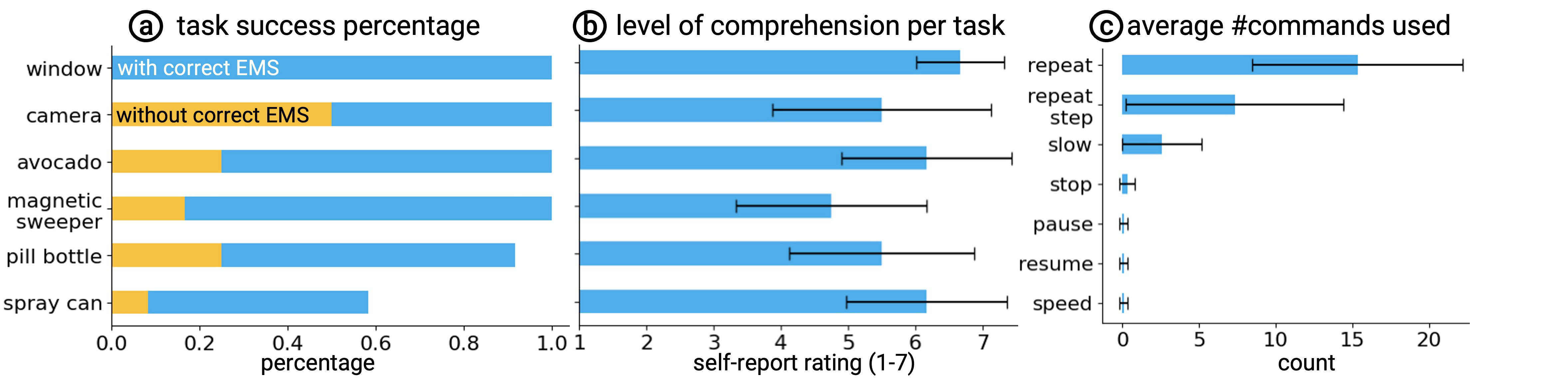}
   \vspace*{-0.4cm}
  \caption{Quantitative results of user study. (a) Task success (\%) (b) Task comprehension (self-report) (c) Usage of verbal commands.}
   \vspace*{-0.4cm}
  \Description{Three bar charts are shown, including task success percentage, level of comprehension per task, average commands used.}
  \label{fig:21}
\end{figure*}

Our study used pre-cached EMS-instructions without any sound output,
i.e., we did not let participants hear the gesture-confirmation in the
study, so that we could observe their reactions to EMS alone. For
this reason, we also set EMS stimulation mode to ``actuate''.

\textbf{Observations}. With participants' prior consent, we filmed their
interactions using cameras and microphones.

\textbf{EMS setup.} Electrical stimulation was delivered via a
medically-compliant stimulator (\emph{HASOMED}, RehaStim) and pre-gelled
electrodes. Prior to the study, the experimenter calibrated the EMS to
ensure pain-free operation and reliable movements. A total of 26
electrodes were placed on a single participant's body: right
\emph{flexor digitorum superficialis}, right and left \emph{extensor
carpi} \emph{ulnaris}, right \emph{deltoid}, right and left \emph{biceps
brachii}, \emph{right extensor digitorum}, right and left \emph{dorsal
interosseous 1\textsuperscript{st}} (as in
\cite{takahashiIncreasingElectricalMuscle2021}), right \emph{flexor
pollicis brevis}, left and right \emph{triceps}, and \emph{splenius
capitis} (as in \cite{tanakaElectricalHeadActuation2022}). We added
electrodes to muscles that were not needed to perform the tasks; these
were used as a \emph{sham} during calibration (i.e., to prevent
participants from inferring the set of needed muscles; thus, we
calibrated more muscles than needed to perform the tasks---e.g., all
tasks could be performed without the \emph{left extensor}).

\vspace*{-0.1cm}
\subsection{Participants}\label{participants}

We recruited 12 participants (five female, six male, and one non-binary; average age=25.3; SD=2.2). All participants were
right-handed. They receive \$12 USD per hour as compensation. Our study
was approved by our ethics review board (IRB23-0025). All participants agreed to video, but some requested to blur
their faces.

\vspace*{-0.1cm}
\subsection{\texorpdfstring{Results of task completion, usage of verbal
commands, and
task-comprehension}{ Results of task completion, usage of verbal commands, and task-comprehension}}\label{results-of-task-completion-usage-of-verbal-commands-and-task-comprehension}

\textbf{Overall success rate}. Overall, in 92\% of all trials,
participants accomplished the task goal by following EMS-instructions
(e.g., took the avocado pit out). Since we utilized two types of
tasks---in one, we first induced a forced-error, and only provided
correct instructions if participants re-prompt (tasks 2-6), and in the
other, we directly provided the correct instructions (task 1)---we break
down this analysis further into its constituents.

\textbf{Success rate when the correct instruction set was delivered
first.} \hyperref[fig:21]{Figure 21} (a) shows that all
participants succeeded in opening a tilt-turn window (which is an
unfamiliar window type in their context). All participants stated they
understood from the EMS-instruction how to turn the handle all the way
up to pull it open, as P10 stated ``{[}stimulation{]} in my shoulder is
telling me to go up (\ldots) until vertical'' (similarly, P3 and P4).

\textbf{Success rate when forced-error instruction set was delivered
first.} For the remainder of the tasks (tasks 2-6), first, the system
delivered a forced-error EMS-instruction set. Still, participants
succeeded in 90\% of the trials. In 28\% of these cases, participants
succeeded without needing to experience the correct instructions (i.e.,
participants succeeded without re-prompting the system in any way).

\begin{figure*}[h]
\centering
  \includegraphics[width=\textwidth, height = 4.4 cm]{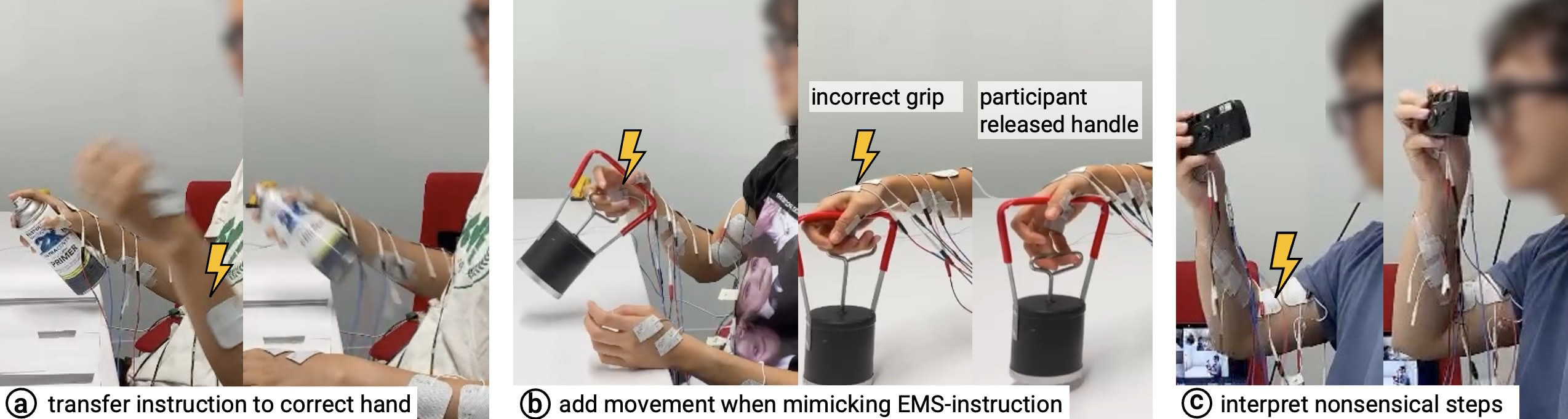}
   \vspace*{-0.6cm}
  \caption{Participants recovered from forced-errors. (a) The participant noticed the instruction was on the wrong hand and transferred the movements to the other. (b) An example of participants adding a release move to recover from the error when mimicking EMS-instructions. (c) A participant making sense of nonsensical elbow movements—looking through the viewfinder.}
  \Description{Photos depicting hands attached with electrodes interacting with various objects, including a spray can, a nail sweeper, and a camera.}
  \label{fig:22}
\end{figure*}

\textbf{Failed trials.} We observed one (out of 72) trial where the
participant failed to meet the task goal---P8 could not, even after
experiencing the EMS assistance, open the pill bottle (which contains a
non-trivial locking mechanism). Additionally, during analysis, we deemed
five spray-can tasks as failed, despite the fact that participants
technically met the goal of painting, but they failed to report
understanding that shaking was required (P1-3, P6, P12).

\textbf{Participants did not take EMS at face-value, they explored
them.} Participants reported they understood the EMS-instructions
(\emph{M}=5.8, \emph{SD}=1.39), as depicted in
\hyperref[fig:21]{Figure 21} (b). We analyzed how participants
interacted with our system by counting the number of commands they used
across all tasks. \hyperref[fig:21]{Figure 21} (c) shows that
participants often used ``repeat'' (\emph{M}=15.3 times,
\emph{SD}=6.87), ``repeat a step'' (\emph{M}=7.3, \emph{SD}=7.11), and
sometimes ``slow down'' (\emph{M}=2.6, \emph{SD}=2.61) commands to
explore EMS-instructions. We further analyzed the correlation between
task comprehension and the number of commands used per task, and found
no significance (\emph{R\textsuperscript{2}}=0.26, \emph{p}=4.42). This
indicates that the participants utilized verbal commands not just for
challenging tasks, but also to explore the system (e.g., after
successfully opening the window with EMS, P5 closed it and said he would
do it again with EMS, visibly exploring our system).

\subsection{\texorpdfstring{Participants' strategy for recovering from
errors and understanding EMS-instructions
}{Participants' strategy for recovering from errors and understanding EMS-instructions }}\label{participants-strategy-for-recovering-from-errors-and-understanding-ems-instructions}

We transcribed participants' comments when thinking aloud and analyzed
their strategies of recovering from erroneous EMS-instructions and how
they understood correct EMS-instructions.

\textbf{Participants recovered without re-prompting.} We observed that
in almost one-third of trials (28\%), participants succeeded by
experiencing the forced-error EMS-instructions and finding their way
around them. We analyzed participants' think-out-loud processes and
found three main strategies of recovering from erroneous instructions.
\textbf{(1) Add movements when mimicking EMS-instructions} (P3, P9, P10,
P12). For instance, P9 found that the magnetic sweeper did not pick up
the nails when EMS made her pull the handle, so she added a release
action on her own to figure out the mechanism. \textbf{(2) Used
partially correct instructions and/or ignored errors} (P1, P3, P4, P9,
P10, P12). After succeeding in the camera task, P3 stated, ``I do not
understand why it also moves the camera {[}shows the wrong gesture{]},
(...) I understand it tells me to move the dial, click to take a
photo''. \textbf{(3) Participants made sense of erroneous steps} (P1,
P4, P6, P11). These participants found new meanings in what we intended
to be a forced-error, for example, P4 interpreted the nonsensical elbow
movements in the camera task as ``it\textquotesingle s trying to let me
look through this window {[}viewfinder{]}'', P6 thought they might be
for ``checking something,'' and P11 described them as ``put camera
down'' and ``put it back''. \textbf{(4) Transfer the instruction to the
correct joint} (P10). In the spray-can task, P10 explicitly pointed out
the error, ignored it, and corrected it, stating, ``EMS demonstrates it
on my left hand, so I\textquotesingle ll do that on my right hand''. See
\hyperref[fig:22]{Figure 22} for study footage of how
participants recovered from forced-errors.

\begin{figure*}[h!]
\centering
  \includegraphics[width=\textwidth, height = 4.4 cm]{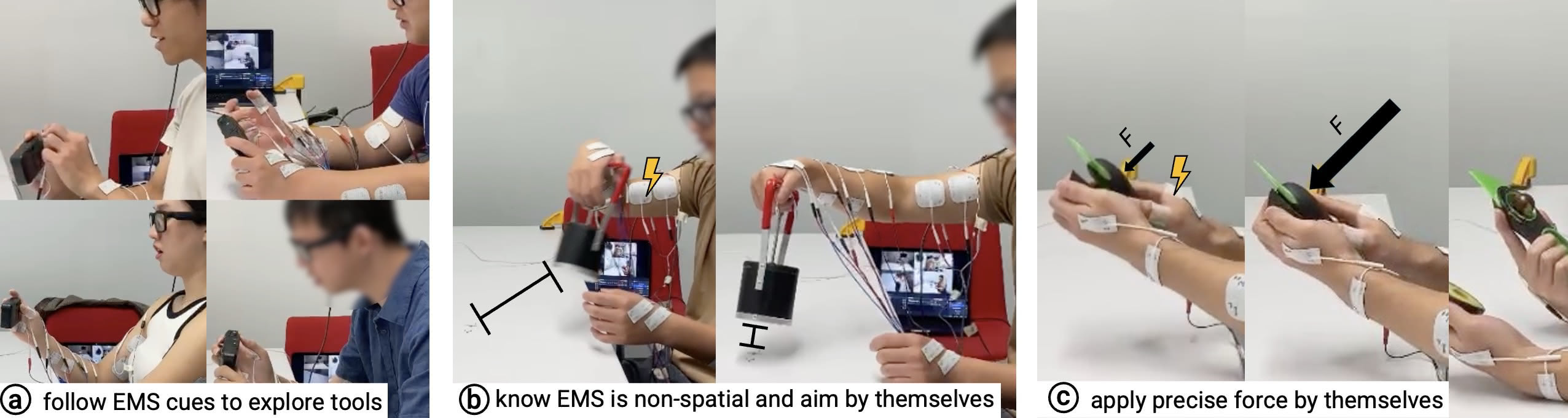}
  \vspace*{-0.7cm}
  \caption{Participants understood correct instructions. (a) Examples of participants following EMS-instructions to discover the mechanism on an analog camera. (b) Participants aiming the nails by themselves as they understood EMS-instructions are non-spatial. (c) Participants adjusted the force on avocado pit as they realized EMS-instructions did not indicate precise forces.}
  \Description{Photos depicting hands attached with electrodes interacting with various objects, including a camera, a nail sweeper, and an avocado cutter.}
  \label{fig:23}
\end{figure*}

\textbf{Participants noticed errors and re-prompted.} In roughly
two-thirds (65\%) of the trials, participants recovered by re-prompting
the system (i.e., saying ``EMS think again''). Before re-prompting, we
observed behaviors that strongly suggest detected errors: \textbf{(1)}
\textbf{Judge if EMS-instructions lead to task outcome} (P1, P2, P4, P8,
P10, P11, P12), participants would let EMS perform the task, then, as
they saw the task was not completed, they would re-prompt. In the
spray-can task, P2 found that the movements in the left hand were just
``shifting the paper around'' and said, ``I am pretty sure this isn't
right (\ldots) EMS, think again.'' P4 tried following EMS-instruction to
turn the pill bottle open, and found ``it's not working, so EMS think
again.'' \textbf{(2) Identify error, re-prompt immediately} (P1, P2, P3,
P5, P6, P7, P8, P12), participants would find a logical mistake or an
unexpected movement for a task, they would instantly re-prompt. For
instance, in the camera task, P2 stated ``EMS repeat second step {[}the
erroneous step{]}, okay, this makes no sense. (\ldots) Maybe EMS, think
again'', or, in the magnetic-sweeper task, after EMS mistakenly
indicated a pull action at the first step, P6 pointed out
``it\textquotesingle s a pull (...) pull this now? (\ldots) no (\ldots)
EMS think again''.

\textbf{Participants' understanding of correct EMS-instructions.}
Finally, we turn our analysis to what participants understood from
experiencing the correct EMS-instructions. We found that, despite never
being extensively explained the limitations of EMS in force-production
or pose-accuracy, they: \textbf{(1) Follow EMS-cues to explore
challenging situations} (P2, P6, P8). Many of our participants
(\textasciitilde25 years old) were not familiar with disposable cameras,
which require rewinding the film after every shot. This made the task
challenging for them. For instance, after experiencing EMS-instructions
for the camera, P2 noticed that ``it {[}the mechanism{]} has something
to do with these two fingers {[}index and thumb{]} (...) Oh, I guess
there\textquotesingle s this little dial here {[}!{]}'' Similarly, P8,
who has never used an analog camera before, asked EMS to repeat the
rewind gesture and found ``There's a wheel over here''. \textbf{(2)
Understood EMS-instructions were non-spatial, adjusted aim} (all 12
participants). For instance, when mimicking EMS-instructions to pick up
the nails with a magnetic sweeper, P12 said, ``Previously, maybe I
didn\textquotesingle t move the tool close enough to the nail, and
that\textquotesingle s why it didn\textquotesingle t stick.''
\textbf{(3) Understood EMS-instructions did not indicate precise force
and adjusted the force} (P6, P8, P12). P8 and P12 depicted canonical
examples of this, with both mentioning that although ``EMS did not
indicate it'', they found the first step of taking the avocado pit out
``requires more force'', which they voluntarily enacted. Similarly, for
the pill bottle (e.g., EMS often does not have the required force to
completely open the mechanism), participants mimicked the
EMS-instruction by themselves with additional force. For instance, P6
succeeded this way. \textbf{(4) Understood EMS-instructions differ from
their own} (P4). It was clear to P4 that EMS' movement trajectories are
not human-like (e.g., EMS movements are much more isolated and typically
use fewer muscle groups). To this end, P4 pointed out he usually uses
the wrist, rather than EMS's choice (elbow), to turn bottle-caps, but
still tried the EMS movement nonetheless to successfully open the cap.
See \hyperref[fig:23]{Figure 23} for study footage of how
participants understood EMS-instructions.

\subsection{\texorpdfstring{User study discussion
}{User study discussion }}\label{user-study-discussion}

\begin{figure}[b]
\centering
\vspace*{-0.2cm}
  \includegraphics[width=\columnwidth]{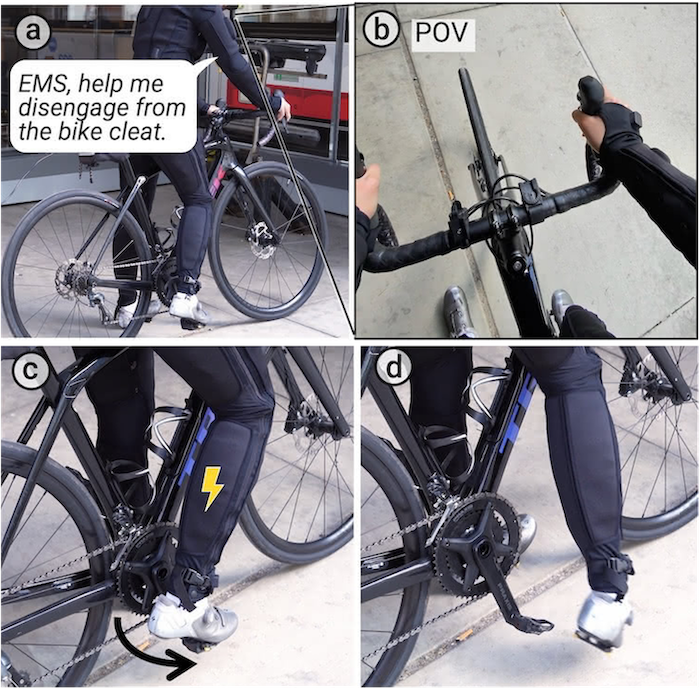}
  \vspace*{-0.6cm}
  \caption{Our system uses (a) user’s spoken request; and (b) POV image, to (c) provide physical assistance through muscle stimulation of her heel, enabling this user to (d) successfully disengage from the bike pedal.}
  \vspace*{-0.2cm}
  \Description{A series of photos depicting a person using our system to disengage from the bike cleat.}
  \label{fig:24}
\end{figure}

We observed how participants utilized EMS-instructions and worked around
system limitations (e.g., hallucinations, EMS imprecision, etc.). We
found that participants not only made use of correct instructions (e.g.,
all participants opened a tilt-turn window) but also detected and
recovered from simulated system errors, inspired by those we observed in
our \hyperref[technical-evaluation-via-an-ablation-study]{\emph{Ablation Study}}. Our study showed the value of providing
procedural knowledge through users' own physical movements---it is
understandable (participants rated 5.8 out of 7 points on how well they
understood the instructions), and it made error-detection easier for the
participants (roughly two-thirds (65\%) of the trials, participants were
able to meet the task goal by detecting errors and re-prompt.) P12
further stated that the ``body's intuitions help notice errors right
away'', whereas text instruction requires reading it and sometimes
referring to a fact-check to detect errors. Participants also noted our
system is helpful for unfamiliar tools (P1, P5, P8), and appreciated its
general-purpose aim~(P5, P12).

\textbf{Study limitations.} Our in lab-study is not without limitations.
To let all participants experience the same set of simulated-errors, we
pre-cached outputs and asked them to assume starting positions. Any
variations outside the study will affect results (e.g., different spoken
requests and poses), so we advise mindful extrapolations of our in-lab
results.

\section{Generate new EMS-assistance with a single
EMS-system}\label{generate-new-ems-assistance-with-a-single-ems-system}

We further demonstrate how our system can provide physical assistance without
task-specific programming. In the example of \hyperref[fig:24]{Figure 24}, a user is practicing cycling
with cycling-shoes for the first time---these lock into the pedals and
require a heel twist to disengage. The user is also wearing our full
system (\emph{Teslasuit} and smart-glasses). To practice safely, she
clips only on her right foot, keeping the other on the ground. As
anticipated, she cannot get her right foot off the clip-pedal, so she
asks, ``EMS help me disengage from the bike cleat'' (\hyperref[fig:24]{Figure 24}). Using contextual clues (e.g.,
POV/pose), our system recognizes that she is wearing
cycling-shoes and the right foot is clipped into the pedal; it infers that the heel needs to be twisted. It confirms by asking, ``I am going to move your right foot''. Upon confirmation, the system proceeds to stimulate her calf (\emph{peroneus longus}) to twist her ankle, disengaging the foot from the pedal.

\begin{figure}[h!]
\centering
\vspace*{-0.1cm}
  \includegraphics[width=\columnwidth]{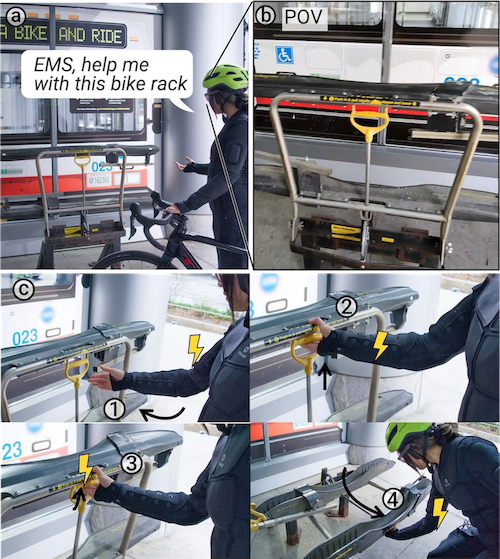}
  \vspace*{-0.6cm}
  \caption{Our system assists a user in placing a bike on a bus rack. (This mechanism is so challenging that the local transportation agency has created an entire life-sized practice location for citizens to practice—where we used our system.)}  
  \Description{A series of photos depicting a person using our system to use a bike rack.}
  \vspace*{-0.2cm}
  \label{fig:25}
\end{figure}

Still feeling like a novice in these cycling-shoes, she decides to take the
bus. However, this is also the first time that she is loading her bike onto the
bus and she is unfamiliar with the bike rack mechanism. Once again, she
asks for assistance: ``EMS help me with this bike rack'' as shown in \hyperref[fig:25]{Figure 25} (a). After
confirmation, the system proceeds to (1) stimulate the biceps to reach the
handle of the bike rack, (2) flexes her hand to grab the handle, (3) squeezes
the fingers to unlatch the rack, and finally (4) contracts the triceps to
lower the rack; similar to our user study, she understands that the EMS
cannot actuate her entire body and completes the remaining
movements, such as bending the legs to help lower the rack (a video version of this example is available at \href{https://embodied-ai.tech}{https://embodied-ai.tech}).

\vspace*{-0.2cm}
\section{Discussion}\label{discussion}

\subsection{Ethics}\label{ethics}

As the first paper that demonstrates a more general-purpose EMS-system,
our work inherits the ethical concerns of both physical assistance
\cite{arnoldMoralDilemmasExploring2017,greenbaumEthicalLegalSocial2016} and AI
\cite{hagendorffEthicsAIEthics2020,kazimHighlevelOverviewAI2021}. We
thus designed our system with these in mind: while it uses involuntary
movements, it still preserves users' autonomy. First, it is
user-initiated---EMS does not activate unless the users invoke. Second,
our \emph{user interactions} module runs in a separate process,
providing always-available control (e.g., users can say ``EMS stop''
when they detect errors, as we observed in the user study). Third, our
system awaits user confirmation before stimulation. Moreover, our system
design followed the established EMS guidelines
\cite{konoDesignStudyMultiChannel2018,omuraElectricalParametersSafe1985,schneegassCreatingUserInterfaces2016}
for safety.

\subsection{Limitations}\label{limitations}

As the first exploration in generating EMS-instructions, our work is not
without limitations. In this section, we discuss them under different
lenses, such as conceptual limitations (e.g., alternative techniques to
implement variants of our proposal, latency, etc.) and practical
limitations (e.g., limited scope of our examples, or the limited
practicality of EMS).

\textbf{Alternative system implementations.} There are many technical
routes to achieve these goals (e.g., reinforcement-learning, advanced
biomechanical simulators). Our key contribution is not to exhaustively
implement all these options but to demonstrate the value of embedding
instruction generation ability in EMS-systems.

\textbf{LLM limitations.} With any system based on LLMs, there is a
chance for ``hallucinations''; while our EMS-knowledge-base and
joint-information constraints will block some errors in instruction
generation, they will affect the selection of stimulation instructions.
At the time of writing, the most advanced cloud-VLM (Gemini) scored a
61.66\% at an image description benchmark, compared to 84.78\% for human
annotators \cite{fossCausalVQAPhysicallyGrounded2025}. Locally-run VLMs scored
lower at 50.06\%. Therefore, our system is limited by the reasoning
capabilities of these VLM to give an accurate contextual description.
Our work relies on OpenAI GPT-4.1, which will likely be surpassed by
newer models. While this presents a limitation of our system, it is also
an advantage---future models are expected to offer higher accuracy
(fewer hallucinations) and reduced latency, both of which constrain our
current approach. In light of this limitation, it is important to
underscore that our contribution is not advancing multimodal-AI but
using it to advance interactive-EMS.

\textbf{Latency.} We run a state-of-the-art model (\textasciitilde175
billion parameters \cite{barretoGenerativeArtificialIntelligence2023})
with general-reasoning capabilities, but it does \emph{not} run locally
nor fast. Accessing it via an internet-API adds latency (e.g., in our
\hyperref[technical-evaluation-via-an-ablation-study]{\emph{Ablation Study}}, the average latency was \textasciitilde23.6s). In
fact, in our video figure, we cached the API responses to skip waiting
on LLM-API calls. Future renditions might run smaller-models locally
(e.g., Llama) in devices with high GPU-parallelism.

\textbf{Example scope.} No range of examples can cover all the ways in
which a user might ask for assistance; our goal is not to cover these
exhaustively but to explore how an EMS-system could \emph{generate}
instructions. Thus, we narrowed down examples to a subset of situations
that (1) require only movement-based assistance (e.g., a user might find
it useful to feel their muscles moving to learn how to operate an
object)---conversely, we did not focus on examples where audiovisual
instructions would have been more beneficial (these are covered by LLMs
already); (2) fit the reasoning capabilities of our LLMs (e.g., do not
need access to hidden information not available in training data or its
sensors); (3) insensitive to our system's latency (e.g., no fast
actions, no need to react timely to external events, and so forth); and,
(4) fit the capabilities of our EMS (e.g., precise-movements are still
considered challenging with state-of-the-art EMS).

\textbf{Practicality of EMS.} Like any interactive system based on EMS,
ours inherits the well-documented limitations found in prior work
\cite{faltaousPerceptionActionReview2022,pfeifferHapticFeedbackWearables2017}.
There are two types of important EMS limitations: those that arise from
the use of electrodes on the skin and those that arise from the
underlying principle of how currents propagate inside the body to
stimulate the muscles. The former includes limitations such as the need
for manual electrode placements, the need for manual calibration, and
``tingling'' sensations at the electrodes. The latter includes the fact
that many muscles are hard to access by means of electrical currents
(e.g., deeper muscles) and that many complex movements that underlie
everyday, manual skills have not yet been demonstrated with EMS. In our
implementation, we explored technical avenues to circumvent some of
these limitations, such as simplifying electrode placement by using an
EMS \emph{Teslasuit} (electrodes placed inside in fixed locations) to
demonstrate applications, and leveraging the state-of-the-art in
electrode layouts (e.g.,
\cite{choudharyAdaptiveElectricalMuscle2025,takahashiIncreasingElectricalMuscle2021})
to achieve the hand movements required for all of our applications and
study. However, as we discuss next, since these limitations are
well-known in EMS research, they are also active areas of research where
progress is underway.

\subsection{Future work}\label{future-work}

The aforementioned limitations are not insurmountable. We discuss how
the landscape of AI and EMS might evolve.

\textbf{Possible improvements on AI reasoning.} First, regarding AI,
just a decade ago, much of the reasoning displayed by contemporary
models was unthinkable. Now, it is expected that newer models will
minimize hallucinations \cite{martinoKnowledgeInjectionCounter2023} and
latency \cite{yangQueueingTheoreticPerspective2024}. Despite projected
advances, some argue that models might never be entirely free of
hallucinations \cite{jiMitigatingLLMHallucination2023}. We expect that
to convert textual-instructions to EMS, future systems might still rely
on concepts from our implementation (e.g., constraining with
EMS-knowledge).

\textbf{Possible improvements on EMS practicality.} Similarly, for EMS,
the capabilities have improved remarkably. Two decades ago, the first
interactive system making use of EMS only actuated a user's biceps
without any interactive closed-loop control \cite{kruijffUsingNeuromuscularElectrical2006}. Just five years
later, researchers demonstrated the interactive systems capable of
actuating movements in smaller joints, such as in the user's wrists or
fingers \cite{tamakiPossessedHandTechniquesControlling2011}. After that,
EMS in HCI rapidly evolved, with improved control-loops (e.g.,
\cite{hwangTelePulseEnhancingTeleoperation2025,kaulFollowForceSteering2016,watanabeFeedbackControlTarget2019}),
independent-finger actuation
\cite{takahashiIncreasingElectricalMuscle2021}, and new body-areas
(e.g., head \cite{tanakaElectricalHeadActuation2022}, legs
\cite{pfeifferCruiseControlPedestrians2015}, and feet
\cite{hassanFootStrikerEMSbasedFoot2017}). With each system, some
technical hurdles were improved, from calibration
\cite{knibbeAutomaticCalibrationHigh2017,usmanFunctionalElectricalStimulation2020},
form-factor \cite{takahashiCanSmartwatchMove2024}, or control-loops
\cite{nithDextrEMSIncreasingDexterity2021}. While there are still
considerable gaps in practicality and accuracy (see \emph{Limitations}),
we expect the processes of electrode placement, calibration, and the
sophistication of control loop algorithms for EMS to improve
significantly in the next decades, which, ultimately, might enable more
complex EMS-movements. Interestingly, as EMS capabilities increase,
programming the stimulation patterns becomes harder. Thus, a system such
as ours that can generate EMS-instructions without
specific-programming might act as a platform to accelerate research.

\textbf{Contrasting EMS with other modalities.} Multimodal-AI has been applied to different interfaces for assistance (e.g., visual and audio interfaces) \cite{arakawaPrISMObserverInterventionAgent2024,huhVid2CoachTransformingHowTo2025,laiLEGOLearningEGOcentric2025,changWorldScribeContextAwareLive2024,leeGazePointARContextAwareMultimodal2024}. Since our contribution
was to endow EMS-systems with the ability to generate their own movement
instructions, we did not study the efficacy of EMS against other
modalities (e.g., visuals, auditory) with respect to physical
assistance. We believe this is an important area for future EMS
research, as it might shine light on where EMS is useful and where it
breaks down for communicating movement instructions. Moreover, as is
typical in multimodal haptics research (e.g.,
\cite{grantAudiohapticFeedbackEnhances2019,marchal-crespoEffectHapticGuidance2013,morrisHapticFeedbackEnhances2007,sigristAugmentedVisualAuditory2013}),
it is very likely that combining EMS with additional modalities (e.g.,
visuals, audio) will yield synergistic effects that might surpass either
modality in isolation.

\textbf{Embodied-AI for skill acquisition}. While we engineered an on-body system that leverages multimodal-AI for flexibly generating muscle instructions for physical assistance, we did not measure whether this new type of embodied-AI can improve physical skill acquisition and "muscle-memory" retention.

\section{\texorpdfstring{Conclusion }{Conclusion }}\label{conclusion}

We engineered an embodied-AI system that assists users with electrical muscle stimulation (EMS)
by delivering muscle-movements akin to the procedural-knowledge needed
to complete a physical task; except, unlike prior EMS-systems, our
assistance is not part of the system's programming. Instead,
instructions are \emph{generated} to \emph{contextually} assist the
user. In this paper, we provide the first instantiation of this novel
concept for EMS, which we investigated via a technical evaluation and a
user study.

Given the rapid evolution in interactive EMS
\cite{faltaousPerceptionActionReview2022,lopesHCIforIntegration}, we believe that such an
architecture will yield a number of benefits for the field of HCI: (1)
it provides a platform for researchers to generate EMS-instructions for
tasks that have never been explored with this type of haptic assistance
before; (2) it can replicate some of the prior work on EMS assistance
(e.g., {[}58{]}) enabling newcomers to this field to explore these ideas
rapidly; (3) it offers a new conceptual way to control EMS-systems---in
our approach, the user controls the system with their voice (re-prompt
it, rewind steps, slow it down---which we observed participants taking
advantage of in our user study). Taken together, we believe that our EMS
architecture enables a new territory for the area of embodied-AI, aiming to inspire
researchers to strive for flexible and context-aware physical assistance. We believe this powers new modes of reasoning with AI that are not just purely audiovisual, but instead focus on delivering \textit{direct} feedback to the user's body. 

\section{Use of AI in this paper}\label{use-of-ai-in-this-paper}

While this paper is about using multimodal-AI to advance EMS-systems, no
AI was used in writing, figures, or ideas.



\begin{acks}

We would like to thank our colleagues: Jas Brooks, for helping with the video; Andre de la Cruz, for the help with user study; finally,  Jasmine Lu and Lonnie Chien, for proofreading the paper. We would also like to extend our gratitude to Teslasuit {\href{https://teslasuit.io/}{(https://teslasuit.io/)}}, especially Andrei Pyko, for providing the EMS-suits used in some of the figures and video (although the reader should note that the presented system is agnostic to the muscle stimulation hardware. In fact, all studies were performed using our custom hardware.)

\textbf{Funding}. This work was supported by multiple funding sources at different times
of our development cycle, including: NSF Grant 2047189, Sony Research
Award Program, University of Chicago Women\textquotesingle s Board, and
the AI Pillar: Human-Machine Creativity at the University of Chicago.
\end{acks}

\balance
\bibliographystyle{ACM-Reference-Format}
\bibliography{reference}

\pagebreak
\appendix

\section{Appendix}\label{appendix}
\subsection{System design comparison}

In this section, we detail situations that exemplify the value of
specific aspects of our architecture (e.g., the value of using body-pose
information to generate movement instructions). Each of these examples
comprises two side-by-side comparisons, visually highlighting the
difference in the output caused by a particular aspect of our
architecture. While these are not essential in judging the accuracy of
our architecture---this was reported in our \hyperref[technical-evaluation-via-an-ablation-study]{\emph{Ablation Study}}---these provide visual examples to assist the reader.

\begin{figure}[h]
\centering
  \includegraphics[width=\columnwidth]{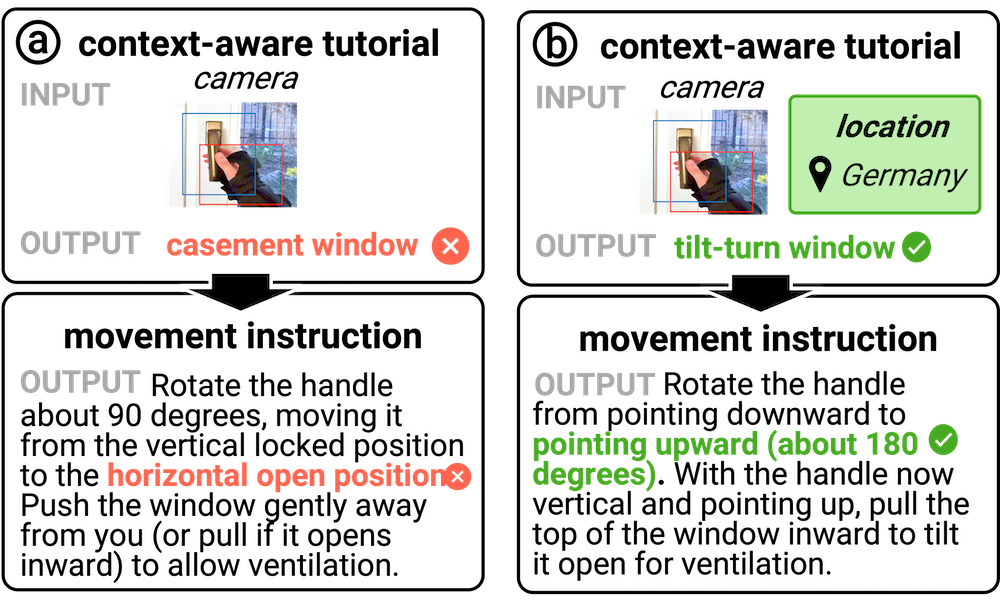}
  \caption{Generated instructions with vs. without location.}
  \Description{Generated instructions with vs. without location. On the left side it is visible (as described in the text) that the LLM did not consider the location of the user and thus failed to recognize the type of window most common in Europe, as shown on right side.}
  \label{fig:26}
\end{figure}

\textbf{Location information.} \hyperref[fig:26]{Figure 26}
depicts side-by-side examples where a sensor alone (e.g., here, GPS
location) enabled an improved outcome (e.g., correctly understanding a
turn-tilt window common in \emph{Europe} vs. assuming a more-standard
casement window).

\begin{figure}[h]
\centering
  \includegraphics[width=\columnwidth]{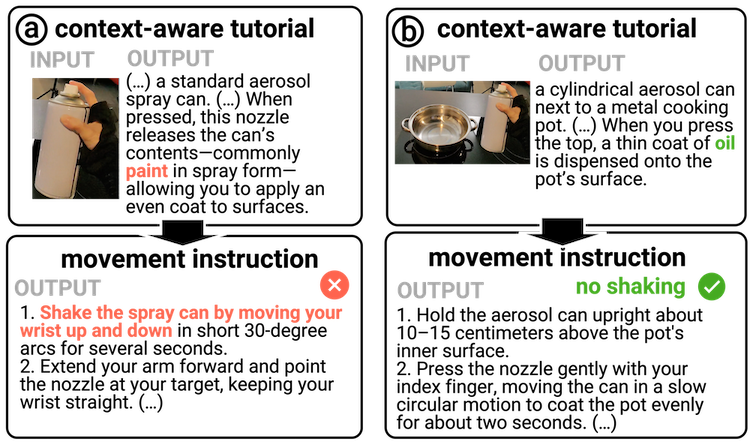}
  \caption{Generated instructions with vs. without POV.}
  \Description{Generated instructions with vs. without POV. On the left side it is visible (as described in the text) that the LLM did not consider the kitchen seen from the user's POV.}
  \label{fig:27}
\end{figure}

\textbf{Visual context.} \hyperref[fig:27]{Figure 27} illustrates
how multimodal-AI can reason in visual scenes. In (a), we cropped the
image of the spray-can and hand, while in (b), we provided the entire
POV. The outcome differs, since without visual-context (e.g., spray and
cooking pan) an incorrect ``shake'' (as in \emph{Affordance++}
\cite{lopesAffordanceAllowingObjects2015}) was generated, though
cooking-sprays do not need shaking.

\begin{figure}[h]
\centering
  \includegraphics[width=\columnwidth]{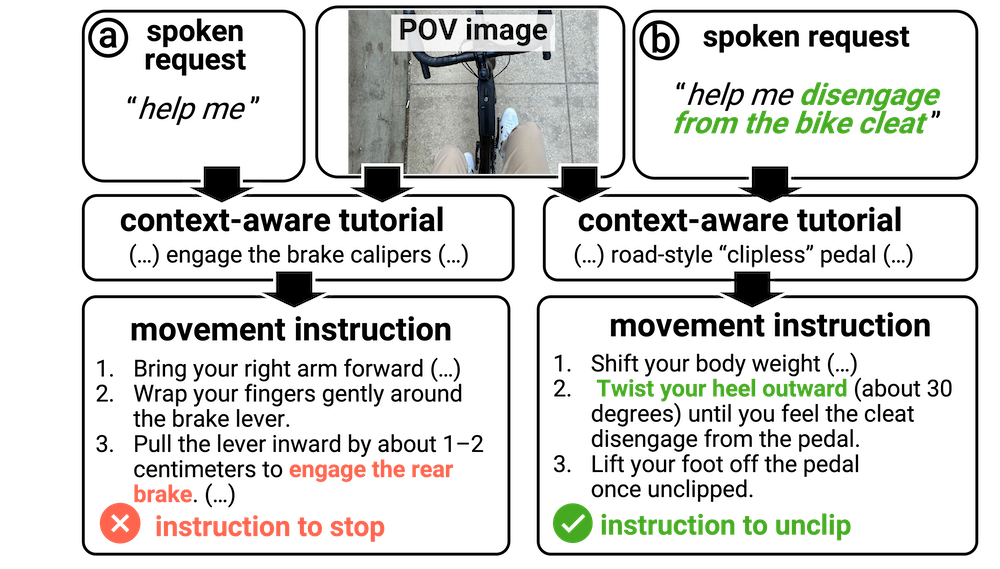}
  \caption{Generated instructions with vs. without details on the user’s request.}
  \Description{Generated instructions with vs. without user request. On the left side it is visible (as described in the text) that the LLM did not consider the user's voice request that mentioned the bike cleat shoes.}
  \label{fig:28}
\end{figure}

\textbf{Spoken request.} \hyperref[fig:28]{Figure 28} illustrates
how user's spoken requests can provide context. In (a), with the POV of
a bike handlebar and a vague ``EMS help me'' request, the system
generated instructions to brake. Conversely, in (b), the addition of
``help me \emph{disengage from the bike cleat}'' provided clues to
generate adequate instructions to unclip the shoe from the pedal.

\begin{figure}[h]
\centering
  \includegraphics[width=\columnwidth]{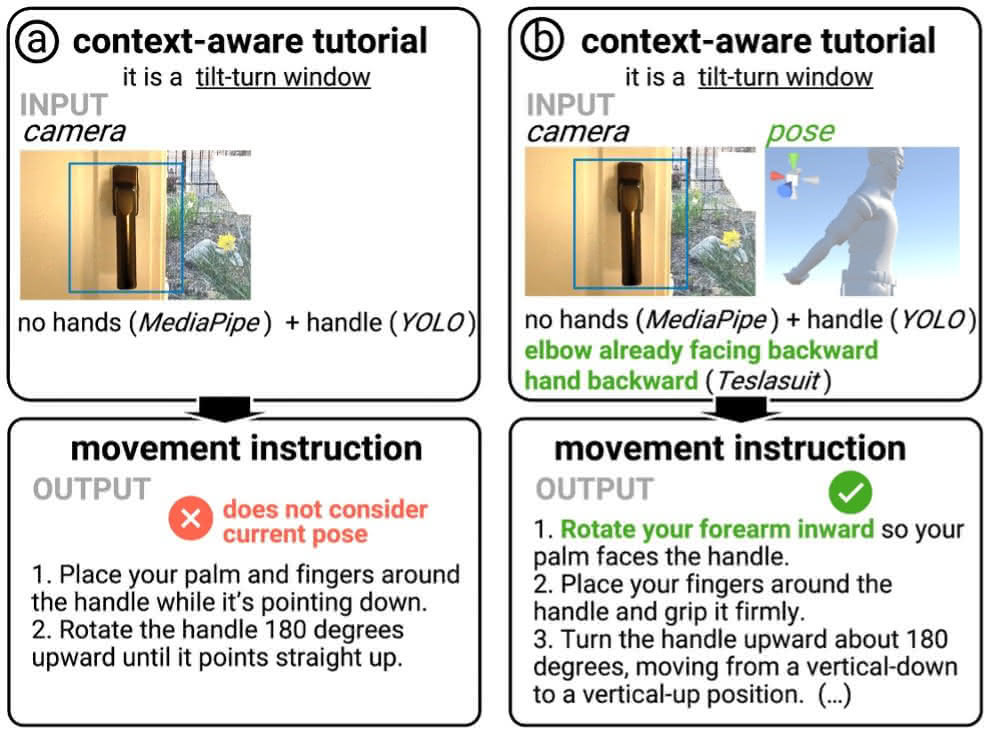}
  \caption{Generated instructions with vs. without body pose.}
  \Description{Image depicts the outputs of two generated instructions with vs. without body pose. On the left side it is visible (as described the text) that the LLM did not consider the current pose.}
  \label{fig:29}
\end{figure}

\textbf{Pose-information.} \hyperref[fig:29]{Figure 29}
illustrates side-by-side examples that exemplify the contribution of
pose-information. In (a) without pose-information, the generated
instructions do not direct the user's hands towards the handle;
conversely, (b), with pose, the generated instructions highlight how the
forearm needs to be rotated to reach the handle, given the current pose.

 Additionally, readers can refer to our open-source repository (\href{https://embodied-ai.tech}{https://embodied-ai.tech}) for the source-code and a database of outputs of our models.








\end{document}